\documentclass[aps,prl,superscriptaddress,twocolumn]{revtex4-2}

\usepackage{amsmath}
\usepackage{amssymb}
\usepackage{xspace}
\usepackage{xcolor}
\usepackage{graphicx}
\usepackage{hyperref}
\usepackage{array}

\newcommand{\plc}{\ell^-}
\newcommand{\PLC}{\mathtt{L}^-}
\newcommand{\flc}{\ell^+}
\newcommand{\FLC}{\mathtt{L}^+}
\newcommand{\stpoint}{(\mathbf{r},t)}
\newcommand{\stpprime}{(\mathbf{r}', t')}
\newcommand{\site}{\mathbf{r}}

\newcommand{\PLCSpace}{\mathcal{L}^-}

\newcommand{\PAlg}{\Sigma_{\PLCSpace}}

\newcommand{\past}{\mathfrak{p}^-}
\newcommand{\Past}{\mathfrak{P}^-}
\newcommand{\future}{\mathfrak{p}^+}
\newcommand{\Future}{\mathfrak{P}^+}

\begin{document}



\title{Unsupervised Discovery of Extreme Weather Events \\Using Universal Representations of Emergent Organization}


\author{Adam Rupe}
\email[]{adam.rupe@pnnl.gov}
\affiliation{Pacific Northwest National Laboratory, Richland WA, USA.}
\affiliation{Center For Nonlinear Studies, Los Alamos National Laboratory, Los Alamos NM, USA.}

\author{Karthik Kashinath}
\affiliation{NVIDIA Corporation, Santa Clara CA, USA.}
\affiliation{NERSC, Lawrence Berkeley National Laboratory, Berkeley CA, USA.}

\author{Nalini Kumar}
\affiliation{Intel Corporation, Santa Clara CA, USA.}


\author{James P. Crutchfield}
\email[]{crutchfield@ucdavis.edu}
\affiliation{Complexity Sciences Center, Department of Physics and Astronomy,
University of California Davis, Davis CA 95616, USA.}


\date{\today}

\begin{abstract}

Spontaneous self-organization is ubiquitous in systems far from thermodynamic
equilibrium. While organized structures that emerge dominate transport
properties, universal representations that identify and describe these key
objects remain elusive. Here, we introduce a theoretically-grounded framework
for describing emergent organization that, via data-driven algorithms, is
constructive in practice. Its building blocks are spacetime lightcones that
embody how information propagates across a system through local interactions.
We show that predictive equivalence classes of lightcones---local causal
states---capture organized behaviors and coherent structures in complex
spatiotemporal systems. Employing an unsupervised physics-informed machine
learning algorithm and a high-performance computing implementation, we
demonstrate automatically discovering coherent structures in two real-world
domain science problems. We show that local causal states identify vortices and
track their power-law decay behavior in two-dimensional fluid turbulence. We
then show how to detect and track familiar extreme weather events---hurricanes
and atmospheric rivers---and discover other novel coherent structures
associated with precipitation extremes in high-resolution climate data at the
grid-cell level.
\end{abstract}

\maketitle

Emergent phenomena are often of primary interest in the study of complex
systems, but disentangling the web of nonlinear interactions that give rise to
them presents enormous challenges to traditional paradigms of scientific
inquiry. Although the broad use of numerical models has significantly improved our ability to study nonlinear systems, difficulties remain. The scale and complexity of model outputs can be such that they are now essentially as difficult to understand as the natural phenomena they approximate. That is, the challenge of uncovering the physical and causal mechanisms of emergent phenomena in a large-scale numerical model can be comparable to doing so in the natural system itself. 

It is well known that emergent \emph{coherent structures} dominate transport in nonequilibrium systems \cite{Hall15a}. While they have clear impacts on transport, these organized structures themselves are often subtle and difficult to identify. A myriad of algorithms have been proposed, based on various \emph{principles of organization}, that attempt to extract coherent structures directly from data. Proper orthogonal decomposition (POD) \cite{Holm12a} and Koopman mode decomposition \cite{Mezi13a} are standard choices for time-independent flow dynamics in the Eulerian frame. Due to these restrictions, however, they are incapable of identifying and tracking localized structures that evolve through time. Lagrangian methods for time-dependent flows are thus preferred, but there are still many competing Lagrangian organizational principles \cite{Hadj17a}. 


The theories underlying most Lagrangian methods are mathematically involved and
the associated data-driven algorithms are costly and do not scale well to large
problems. In addition, their organizational principles may not always capture
the structures of interest to domain scientists. This results in a disconnect
between developing general principles of organization and deploying data-driven
algorithms in practice by domain scientists striving to detect coherent
structures. Due to this, domain practitioners fall back on automated heuristics
that are specialized to domain-specific problems. 

As a prime example, consider the Earth system---a quintessential complex system
\cite{ghil20a,Ravi22a}. The emergent behaviors associated with climate change
present one of the most pressing issues of our time \cite{ipcc22a}. Many of the
direct impacts of climate change on society are felt through emergent
organization in the form of \emph{extreme weather events} \cite{tipp18a}, such
as hurricanes and atmospheric rivers (AR). Moreover, as the Earth warms, the
character of localized extreme weather events (EWE) has and will continue to
change \cite{Rob20a,robi21a}. 

Global climate models are the primary tool used by scientists to study EWEs and
climate change more generally \cite{edwa10a,ipcc13a}. While high-resolution
models produce EWEs under a variety of warming scenarios, discovering the underlying mechanisms driving specific shifts in intensity, duration, and
spatial dynamics of individual events is under active investigation by researchers. Large-scale surveys of various models
and warming scenarios \cite{eyri16a} are thus essential to study statistical changes
in EWE behaviors, but detailed assessment is made difficult due to the sheer
quantity of data produced.
For example, CMIP 6 is estimated to contain 15-30 PB of total climate data
\cite{stoc17a}. There is an immediate need in the climate community for
automated discovery and tracking of EWEs to better understand and predict how
they are changing in a warming world and to help mitigate their deleterious
impacts. Discovering EWEs in a principled, robust, and scalable manner remains
an outstanding challenge. 

Current practice in the climate community employs automated heuristics based on
meteorological properties (e.g., temperature and pressure thresholds) to
identify EWEs from data \cite{Neu13a,Shie18a}. In addition to being somewhat ad
hoc, these heuristics are specialized to particular types of structure they
identify. The heuristics used to identify hurricanes, for example, differ from
those used to identify ARs. More worrying, hurricane heuristics used in
one climate model may differ from those used in another---with, say, higher
resolution. At best, this makes comparing EWE behavior across climate change
scenarios challenging. In addition, the changing climate itself may lead to
qualitatively different EWE behavior that eludes current heuristics. 

In machine learning terminology, identifying individual events at the
single-pixel or model-grid-cell scale is known as a \emph{segmentation}
analysis of extreme weather events. At first glance, EWE detection and tracking
may seem an ideal use case for deep learning algorithms that have brought
dramatic advances in computer vision. Indeed, there is considerable effort
devoted to extreme weather segmentation using deep learning
\cite{mudi17a,jian18a,cohe19a,kurt18a}. 

Deep learning's main weakness, though, is that neural networks have been most
successful at \emph{supervised} computer vision tasks. In this, the networks
are trained on \emph{ground truth} labeled examples---data exemplars typically
labeled by human experts. As with general fluid vortices \cite{Epps17a}, no
ground truth is available for extreme weather events; see, e.g., Refs.
\cite{Neu13a,Shie18a}. More to the point, \emph{there is no generally agreed
upon objective definition of an extreme weather event.} They are currently
identified on a ``know it when you see it'' basis. For supervised deep
learning, the automated heuristics implemented in the TECA code
base~\cite{Prab12a} are used as surrogate ``ground-truth'' training examples,
but the neural networks' high evaluation scores show that they merely learn to
reproduce the output of these heuristics. More recent efforts have turned to
curated collections of expert-labeled images for training \cite{prab21a}. While
more promising for practical application, this approach relies heavily on
subjective human evaluation and not on a general organizational principle.

The absence of objective definitions of Lagrangian coherent structures and
extreme weather events are symptomatic of a more general deficiency:
\emph{there is no objective definition of emergent organization}. Therefore,
there is no objective way to evaluate any proposed organizational principle and
associated unsupervised algorithms that discover structures as instances
satisfying the principle. The current recourse is to rely on subjective, and
largely visual, intercomparison among competing principles \cite{Hadj17a}. In
addition, there is a larger question of whether or not there is a \emph{single}
principle that can capture the wide array of organized structures produced in
complex spatially-extended dynamical systems. 

The following answers this question by proposing \emph{the local causal states
as universal representations for emergent organization}. With foundations in
complex systems theory and the physics of self-organization
\cite{Crut12a,rupe22a}, the local causal states decompose a system's behavior
into minimal, causally-interacting components. An organizational principle
using (deviations from) hidden spacetime symmetries revealed by the local
causal states is able to extract coherent structures in cellular automata
models \cite{Rupe18a} and Lagrangian coherent structures in complex fluid flows
\cite{Rupe19a}.

Here, we report on two contributions towards a universal and actionable organizational
principle using the local causal states: (i) A quantitative verification method
for coherent structure discovery that identifies the nonstationary power-law
relaxation of vortex dynamics in two-dimensional free-decay turbulence. And,
(ii) 
a 
global extreme weather event
segmentation in high-resolution general circulation model (GCM) data. Beyond
known EWEs---such as, hurricanes and atmospheric rivers---we demonstrate that
the local causal states extract novel structures associated with precipitation
extremes.

\subsection{Local Causal States}
We now introduce our organizational principle and compare it with two other
general approaches to structure detection---proper orthogonal decomposition and
finite-time coherent sets. True to their name, the \emph{local causal states}
are defined using a weak notion of causality. Not only does the future follow
from the past, but in spatiotemporal systems that evolve through local
interactions, there is a limit on how fast causal influence can propagate. This
limit defines \emph{lightcones} in the system that are essential features used
in constructing local causal states. The \emph{past lightcone} ($\PLC$) of a
point in spacetime is the collection of all points at previous times that could
possibly have influenced the spacetime point through the local interactions.
Similarly, the \emph{future lightcone} ($\FLC$) of a spacetime point is the
collection of all points at later times that the spacetime point could
influence through local interactions.

 \begin{figure}
\begin{center}
\includegraphics[width=1.0\columnwidth, trim={1.25cm 0cm 3.8cm 0.8cm},clip]{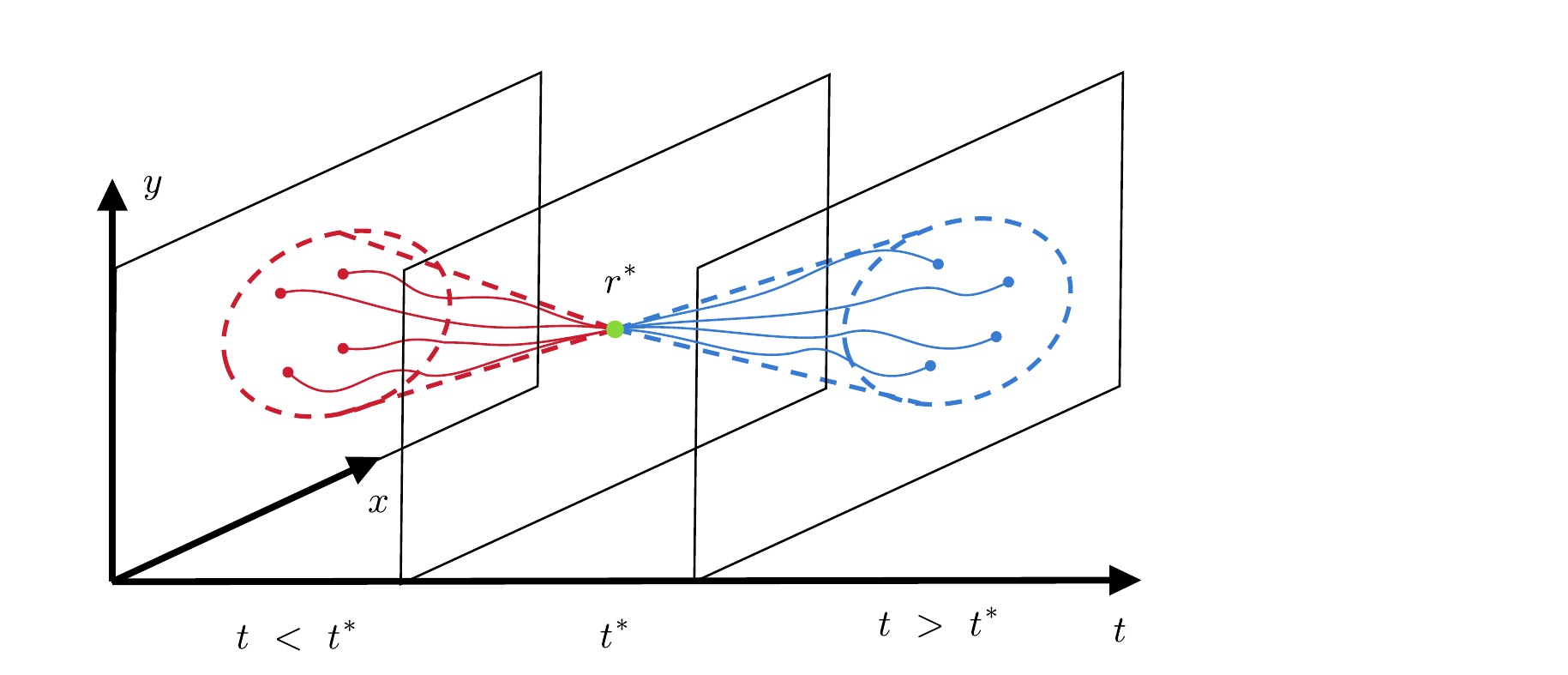}
\end{center}
\caption{Co-occurring past (red) and future (blue) Lagrangian lightcones at spacetime point $(r^*, t^*)$ shown with dashed lines. All possible Lagrangian trajectories (examples shown by solid lines) leading to and emanating from $(r^*, t^*)$ are contained within the lightcones.
	}
\label{fig:lightcones}
\end{figure}

Local causal states are then defined through the \emph{local causal equivalence
relation}: two past lightcone configurations, denoted $\plc$, are considered
causally equivalent, denoted $\sim_\epsilon$, if they have the same
distribution over co-occurring future lightcones (see Figure~\ref{fig:lightcones}):
\begin{align*}
    \plc_i \sim_\epsilon \plc_j &\iff \Pr(\FLC | \PLC=\plc_i) = \Pr(\FLC | \PLC=\plc_j)
    ~.
\end{align*}
The local causal states are the equivalence classes of the local causal
equivalence relation. A given local causal state is a set of past lightcone
configurations that have the same distribution over co-occurring future
lightcones \cite{Rupe18a}. This is expressed functionally by the
map $\epsilon(\cdot)$ from past lightcone configurations to a local
causal state: $\epsilon(\plc_i) = \epsilon(\plc_j)$ if and only if $\plc_i$ and
$\plc_j$ belong to the same local causal state. 

\begin{figure*}
\begin{center}
\hspace*{0cm}\includegraphics[width=1.0 \textwidth]{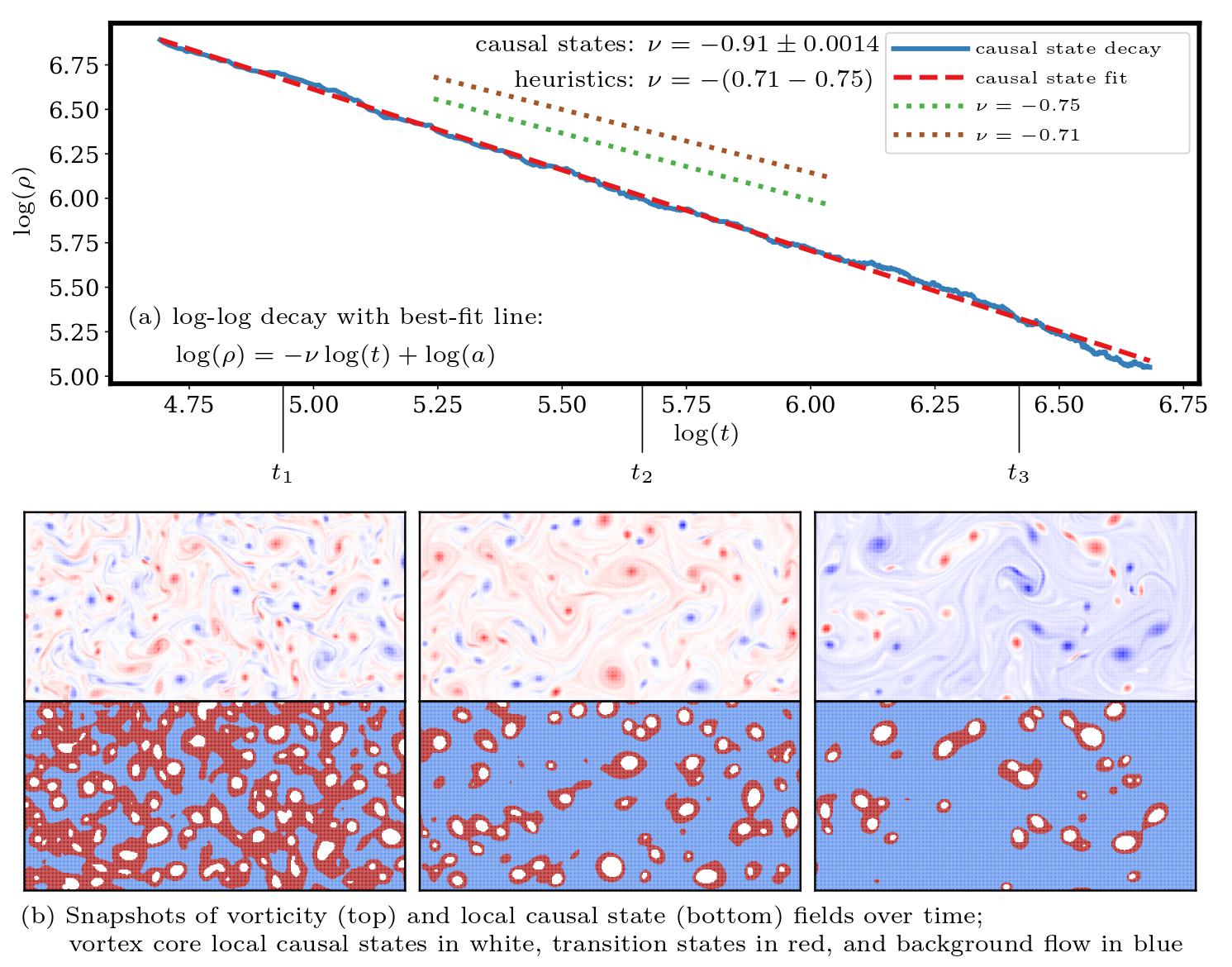}
\end{center}
\caption{Local causal state vortex cores over time and their power law decay. }
\label{fig:vortex_decays}
\end{figure*}

Crucially, the $\epsilon$-function provides a pointwise mapping from points in
spacetime to corresponding local causal states, through their past lightcones.
When applied to all points in a spacetime field $X(\mathbf{r}, t)$ this
produces a corresponding \emph{local causal state field}, denoted
$S(\mathbf{r}, t) = \epsilon\bigl(X(\mathbf{r}, t)\bigr)$. Since
$\epsilon(\cdot)$ is a pointwise mapping, applying at each point in spacetime,
the resulting local causal state field $S(\mathbf{r}, t)$ shares the same
coordinate geometry as its associated spacetime field $X(\mathbf{r}, t)$.

This means the local causal states provide a \emph{spacetime segmentation}, as
desired. That is, the finite set $\mathcal{S} = \{S_i\}$ of local causal states can be thought of as a set of \texttt{class labels} $S_i$ and each
point in spacetime is assigned one of these local causal state labels through
the $\epsilon$-function. Since the $\epsilon$-function is a local mapping
defined using lightcones, it is \emph{equivariant} under spacetime
translations, rotations, and reflections. This guarantees that the local causal
state representations do not depend on spacetime location or orientation of structures they extract. 

Data-driven approximation of the $\epsilon$-function is achieved using two
stages of clustering \cite{Rupe19a}. The first clustering stage $\gamma$ is a
distance-based partitioning of the space of finite-depth past lightcones, which
we implement using the K-Means algorithm \cite{kmeans}. The $\gamma$-function
maps past lightcones to their associated distance-based cluster. Performing
distance-based clustering on past and future lightcones creates sets of cluster
labels so that the distributions $\Pr\bigl(\gamma(\FLC) | \gamma(\plc)\bigr)$
can be empirically approximated through simple counting. During the second
clustering stage $\psi$, elements of the $\gamma$-partition are clustered
together if they have approximately the same empirical future-lightcone
distribution $\Pr\bigl(\gamma(\FLC) | \gamma(\plc)\bigr)$. Elements of the
$\gamma$-partition are mapped to their resulting clusters under the
$\psi$-function. The approximated $\epsilon$-function used for inference is
thus given as:
\begin{align*}
    \epsilon(\plc) \approx \psi \bigl(\gamma(\plc)\bigr)
    ~.
\end{align*}

Systems under study often have multiple physical fields of interest; in the
climate domain these include temperature, pressure, wind speeds, and water
vapor. In these multivariate cases, the spacetime fields and lightcones
extracted from them are tensors, with each point in spacetime having a vector
of values over the multiple physical fields. Multivariate local causal state
analysis is performed using a tensor lightcone metric that computes distances
for the $\gamma$-partition using values from multiple physical fields. Using
this tensor lightcone metric, the local causal states may be tuned to the
specific physics of the system. See the Supplementary Materials for more
details on the local causal states and their data-driven reconstruction. 

At present, the canonical representation learning method for spatiotemporal
systems is the proper orthogonal decomposition (POD) \cite{Holm12a}. In this, a
fixed set of spatial modes are found that provide an optimal reconstruction of
the spatiotemporal system through linear superposition using time-varying
coefficients. 

POD modes, also called Empirical Orthogonal Functions in the climate literature, are the learned representations from which we may extract potential
coherent structures. The local causal states can be seen as a representation
learning method similar to POD, in that it learns a finite set of ``template''
representations. However, rather than learning full spatial field templates
(modes), the local causal state templates are localized and assigned at each
point in space and time. This greatly increases their representational
capacity compared to POD modes, as they can be arranged arbitrarily in space and time. 

Since POD modes are spatially global (Eulerian) and fixed in time, they are
structurally very limiting and are incapable, for instance, of tracking the
evolution of spatially-localized structures over time. Lagrangian methods are
thus better suited for coherent structures that dominate material transport in
fluid flows. The Lagrangian approach most similar to the local causal states
are the \emph{finite-time coherent sets} \cite{froy07a,froy10a} that identify
contiguous regions in space at a given time such that the points in that region
mostly evolve together (coherently) under the Lagrangian flow. Most simply,
they are estimated as eigenfunctions of the time-dependent Perron-Frobenius
operator (defined from the Lagrangian flow map) with eigenvalues close to
unity. 

Noting the importance of Lagrangian dynamics for coherent structures in fluids,
we use \emph{Lagrangian lightcones} when constructing local causal states
of fluid flows. Rather than tracking information propagated through local
interactions, Lagrangian lightcones track information propagated through
advection. Specifically, all possible Lagrangian trajectories that could lead
to (from) a point in spacetime are contained within its past (future)
Lagrangian lightcone, as shown in Figure~\ref{fig:lightcones}.

The collectively-evolving bundles of trajectories identified as coherent sets
produce a distinct signature in the lightcones of points in the set. Therefore
coherent sets correspond to a distinct set of local causal states. Similarly,
the boundaries of coherent sets are hyperbolic Lagrangian coherent structures
that act as transport barriers. As observed in Ref.~\cite{Rupe19a}, hyperbolic
Lagrangian coherent structures also correspond to boundaries of local causal
states. The transport barrier produces distinct lightcone signatures on either
side, resulting in two different local causal states across the boundary. 



\subsection{Vortex Decay in Two-Dimensional Turbulence}
Lacking an objective definition of coherent structures, how does one evaluate
proposed organizational principles and their ability to identify structures
from data? As mentioned above, qualitative visual comparison is typical. In
Ref.~\cite{Rupe19a} local causal states are visually compared to other
Lagrangian methods assessed in Ref.~\cite{Hadj17a}, with good agreement between
the local causal states and the leading Lagrangian methods for identifying
vortices specifically. Here, though, we go further to \emph{quantitatively}
evaluate coherent structure identification using the local causal states'
ability to capture nonstationary behavior, including coherent structures with
finite lifespans.

We demonstrate this by analyzing a longstanding problem in
hydrodynamics---vortex detection in two-dimensional free-decay turbulence
\cite{mcwil90a}. Like-signed vortices---the coherent structures of
interest---undergo pairwise annihilation. This results in a nonstationary decay
behavior with a power-law decay rate:
\begin{align}
\rho(t) = a\;t^{-\nu}
~,
\label{eq:power_law}
\end{align}
where $\rho(t)$ is the number of vortices and $\nu$ is the vortex
decay rate. Empirically, $\nu$ is observed to be $\nu \sim 0.71-0.75$
\cite{mcwil90a,carn91a}. Extracting the decay rate provides an opportunity for
a quantitative comparison with known physical behavior. 

Recall that domain scientists employ heuristics specialized to their particular problem, rather than general organizational principles, to identify their structures of interest. The power-law decay and the range of accepted decay rates in two-dimensional turbulence are identified using tailored vortex heuristics based on vorticity thresholding along with geometric considerations \cite{mcwil90a}. 

The local causal states are able to identify vortices in the two-dimensional turbulent flow using a general organizational principle: \emph{coherent structures are spatially-localized and temporally-persistent deviations from spacetime symmetries in the local causal state field} \cite{Rupe18a}.
Local causal state segmentation distinguishes between vortices spinning in
different directions, but estimating the decay rate requires tracking only the
total number of vortices. Therefore, segmentation is performed on the absolute
value of vorticity, resulting in three local causal state class labels, as Figure~\ref{fig:vortex_decays}
(b) shows. These labels can be qualitatively interpreted as the background
potential flow (blue), vortex cores (white), and transition regions (red)
surrounding the vortex cores \cite{mcwil90a}. The background potential flow state acts as the hidden spacetime symmetry that is broken locally by the vortex states.

Quantitative analysis comes from algorithmically counting the number of
vortex cores at each time using a union-find algorithm \cite{fior96a}.  A
log-log plot of the number of vortex cores over time is shown in
Figure~\ref{fig:vortex_decays} (a). Since the decay rate is a power law, the
log-log plot is linear, with the decay rate $\nu$ given as the slope. 
Note though that we fit the power law in Eq.~(\ref{eq:power_law}) directly using \texttt{scipy.optimize.curvefit}; a log-log plot is shown in Figure~\ref{fig:vortex_decays} (a) for visual clarity.

The (log-log) best-fit line shows the decay rate of local causal state vortex cores over time
fits very well to a power law, with decay rate $\nu \approx 0.91$. Slopes
representing the accepted empirical range $\nu = - (071-0.75)$ are shown for
comparison. As seen in the full spacetime segmentation video \cite{turbseg},
the higher-than-expected decay rate is due to a small number of vortex core
states eventually being mislabeled as background, and so they vanish without
pairwise merging. However, the overall behavior is dominated by the correct
physical mechanism of pairwise vortex merging, giving a strong fit to the power
law decay.

We note that finite-time coherent sets are the only other Lagrangian method
able to capture the nonstationary birth and death of individual structures over time
\cite{froy21a}. However, the required mathematics is quite complicated.
Whereas, the ephemeral behavior of structures with finite lifespans is
naturally encapsulated by the local causal states without additional
modification. It remains to be seen if the Lagrangian finite-time coherent set
algorithm can scale to larger problems and reproduce quantitative relationships
like the vortex decay just described.

\begin{figure*}[t]
\begin{center}
\includegraphics[width=0.9 \textwidth]{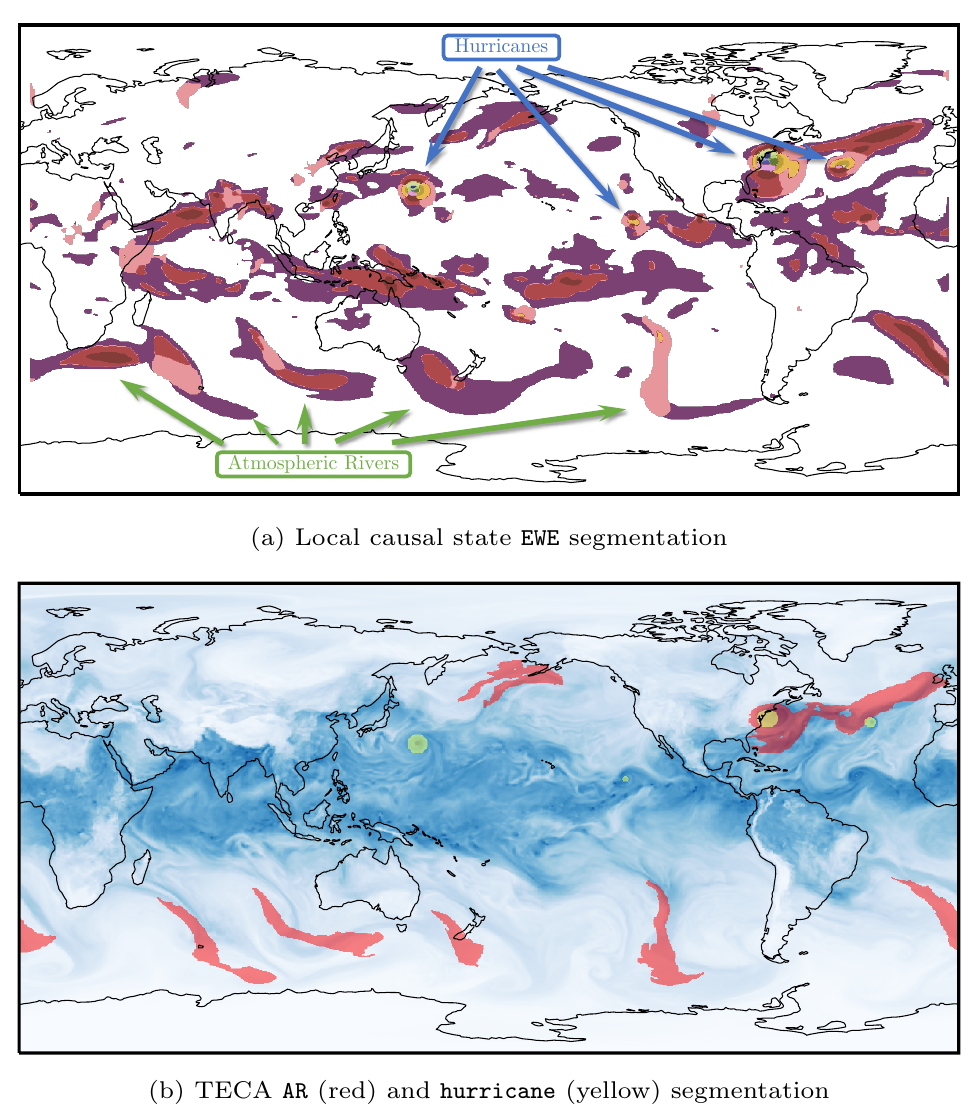}
\end{center}
\caption{Hurricane segmentation masks created from local causal state segmentation of the IVT field.
	}
\label{fig:EWE-seg}
\end{figure*}

\subsection{Discovering and Tracking Extreme Weather Events}

Employing data from the 0.25-degree CAM5.1 Global Circulation Model
\cite{wehn14a} the following analyzes extreme weather events in global climate
data. The goal of our analysis is twofold: (i) simultaneously identify known structures, such
as hurricanes and ARs, using a general and robust organizational principle (rather than specialized heuristics or supervised deep learning), and (ii)
explore as-of-yet undiscovered structures and their relevance for precipitation
extremes. The physics of EWEs is incorporated into the local causal states
through multivariate segmentation using wind velocities and column-integrated
water vapor. This approach is based on the integrated vapor transport (IVT)
field, commonly used by climate scientists to study EWEs \cite{sous20a}.

Performing a univariate segmentation on the scalar IVT field itself, we find a set of
local causal states that visually coincide with hurricanes. That is, local
causal states in this \texttt{hurricane} set (almost) only occur in spacetime
locations where a hurricane is present. Using the shared coordinate geometry,
we can then create an unsupervised hurricane tracker by simply highlighting
spacetime points where these \texttt{hurricane} states occur.
A snapshot of this tracker and its construction is shown in Figure~\ref{fig:ivt_hurricane_mask} in the Supplementary Information.
The full spacetime video of this hurricane tracker can be
seen in~\cite{hurtrac}. 

Atmospheric rivers are more difficult to isolate with a unique set of \texttt{AR} states from a similar IVT field segmentation. However, the more general extreme weather segmentation we now describe does encompass ARs, along with other EWE coherent structures.

In contrast to conforming the local causal state analysis with known EWE
structures as above (which, recall, are not objectively defined), we now turn
to novel EWE structure discovery. Following the local causal state coherence
principle of localized structures breaking a hidden spacetime symmetry, we seek
to identify sets of local causal states breaking a uniform background state,
similar to the blue background state for turbulent vortices shown in
Figure~\ref{fig:vortex_decays} (b). Such a background state is not recovered
using a univariate segmentation of the IVT field nor the integrated vapor
field; see Figure~\ref{fig:ivt_hurricane_mask}~(a). 

A uniform background symmetry state is recovered, however, using a multivariate
segmentation with the tensor lightcone metric given by
Equation~(\ref{eq:ivt_alt_metric}) in the Supplementary Information. Motivated by the construction of the IVT field, this multivariate segmentation utilizes the column-integrated water vapor field, and the two component fields of mid-column wind velocity. A snapshot
of the local causal state field from this multivariate segmentation is shown in
Figure~\ref{fig:EWE-seg} (a). Video of a general EWE tracker, similar to the
hurricane tracker above, using these local causal states is seen in
\cite{ewetrac}. The uniform background local causal state is colored white,
with all other colors then corresponding to coherent structures identified by
the local causal state definition. 

We label the nonbackground local causal states as \texttt{EWE} states since they simultaneously encompass hurricanes and atmospheric rivers. To justify this statement, we include a comparison with the TECA hurricane \cite{Prab12a} and AR \cite{Obrien20a} heuristics on the same dataset, shown in Figure~\ref{fig:EWE-seg} (b) overlaid on top of the integrated water vapor field. TECA and the similar TempestExtremes \cite{Ullri21a} are the state-of-the-art software packages that implement extreme weather segmentation heuristics used in practice by climate scientists. 

Visually, we can see in Figure~\ref{fig:EWE-seg} that \texttt{hurricane} (yellow) and \texttt{AR} (red) structures identified by TECA in (b) are encompassed by \texttt{EWE} local causal states in (a). Quantitatively, $82\%$ of spacetime points in the dataset identified as either \texttt{hurricane} or \texttt{AR} by TECA are also identified as \texttt{EWE} by the local causal states. Recall that TECA and related heuristics are \emph{not} ground-truth for these structures, and so it is not necessarily desirable for the local causal states to cover $100\%$ of the TECA segmentation points. In fact, in our dataset the TECA AR heuristics have a false positive signal that identifies a hurricane as an AR, as confirmed by the author of the TECA AR heuristics in personal communication; see Figure~\ref{fig:teca-false} in the Supplementary Information. 

In addition to the known hurricane and AR structures, there are many additional
coherent structure signatures in the local causal state segmentation that do
not correspond to known extreme weather events. To test their significance, we
investigate the co-occurrence of precipitation extremes \cite{Catto13a} with
these \texttt{EWE} local causal states. That is, how many precipitation
extremes occur in a spacetime location for which that same location in the
local causal state field is an \texttt{EWE} state, rather than a
\texttt{background} state?

\begin{table}[h!]
\centering
    \begin{tabular}{ | m{2.0cm} || >{\centering\arraybackslash} m{1.8cm}| >{\centering\arraybackslash} m{1.8cm}| } 
        \hline
          & LCS & TECA \\
         \hline 
         \multicolumn{3}{|c|}{$90^{\textrm{th}}$ percentile extremes}\\
         \hline 
         global & 45.0\% & 11.0\%  \\
         tropics & 61.7\% & 0.5\% \\ 
         extra-tropics & 40.8\% & 23.3\% \\
         \hline
         \multicolumn{3}{|c|}{$99^{\textrm{th}}$ percentile extremes}\\
         \hline
         global & 58.1\% & 27.3\% \\
         tropics & 69.5\% & 1.43\% \\ 
         extra-tropics & 57.8\% & 43.3\% \\
         \hline
         \multicolumn{3}{|c|}{$99.9^{\textrm{th}}$ percentile extremes}\\
         \hline 
         global & 75.3\% & 21.8\% \\  
         tropics & 76.7\% & 2.73\% \\ 
         extra-tropics & 77.8\% & 51.3\% \\
         \hline
    \end{tabular}
\caption{Percent of extreme precipitation events co-occurring with local causal state (LCS) and TECA segmentations.}
\label{tab:extremes}
\end{table}

Precipitation extreme results from the full three month time span of our
climate dataset are shown in Table~\ref{tab:extremes}, with analogous TECA
results given for comparison. Three different percentile thresholds are given
for defining the extremes, and for each case results are given globally, as
well as broken down between tropics and extra-tropics. The zonal breakdown is
given because the hurricane and AR structures identified by TECA are largely
extratropical phenomena. (The TECA AR heuristics explicitly filter out
potential signals in the tropics.) In all cases, the LCS \texttt{EWE} states
capture a significantly larger proportion of precipitation extremes than the
combined TECA segmentation. Given that $82\%$ of the TECA  segmentation is
identified as \texttt{EWE} by the local causal states, this means the novel
\texttt{EWE} coherent structures identified by the local causal states are
significantly associated with precipitation extremes. Thus, we identify these
coherent structures as novel extreme weather events discovered by the local
causal states. Note that both the local causal state and TECA segmentations are
created without any direct use of precipitation fields.

\begin{figure*}[t]
\begin{center}
\includegraphics[width=1.0 \textwidth]{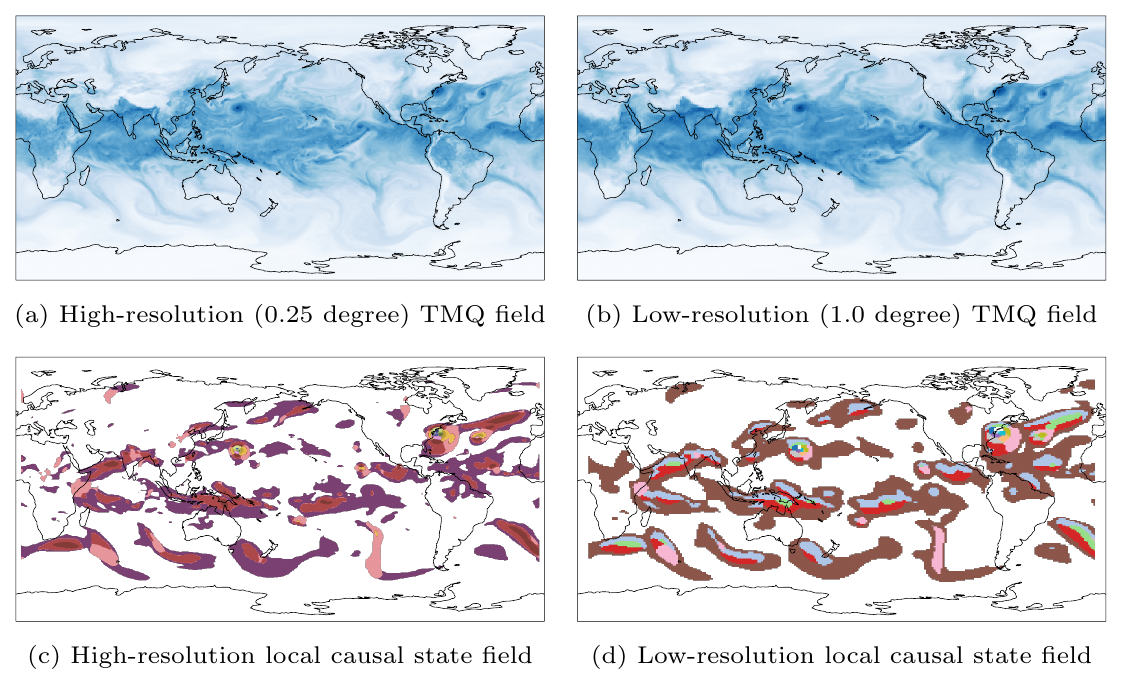}
\end{center}
\caption{Snapshots of water vapor field from high- (a) and low-resolution (b) versions of our CAM5.1 data, with the corresponding local causal state field snapshots shown in (c) and (d), respectively. 
	}
\label{fig:multi-res}
\end{figure*}

To close, we emphasize that
the standard extreme weather segmentation paradigm \cite{Hodges94a} used
by TECA and TempestExtremes heuristics are in the Eulerian framework, unlike
the Lagrangian local causal states. The known shortcomings of Eulerian coherent
structure segmentation \cite{Hadj17a} contribute, at least in part, to the
inability of extreme weather heuristics to generalize across climate data. 
Crucially, hurricane and AR heuristics are not robust to the resolution of
climate models \cite{LiF13a}. In contrast, the general organization principle
embodied by the local causal states is robust across model resolution, as demonstrated in Figure~\ref{fig:multi-res}. We employed a 4x4 block-averaging (mean pooling) filter to reduced the resolution of our climate dataset fourfold, from $0.25$ degree resolution to $1.0$ degree. A local causal state segmentation was performed on this low-resolution data with the same parameters used in the high-resolution segmentation (except for the ``speed of light'' which is reduced because we are changing spatial resolution in the dataset without changing temporal resolution). 

Visually, the resulting local causal state field is very similar to that from
the high-resolution data. However, we note that the two segmentations and their
local causal states are distinct and should not be directly identified. The 4x4
block averaging operation is incommensurate with the odd-length dimensions of
the lightcone spatial slices. Thus, there is not a direct relation between the
lightcones of the high- and low-resolution datasets, nor between their local
causal states. Hence, we have kept distinct color labels of unique local causal
states (assigned randomly by our algorithm) in Figure~\ref{fig:multi-res} (c)
and (d). Importantly, though, the \emph{segmentation semantics} of
\texttt{background} and \texttt{EWE} \emph{are} shared between the local causal states across resolutions.

\subsection{Conclusions}
Advances in dynamical systems and applied ergodic theory are forming the
foundations of a new paradigm for data-driven science \cite{berr20a}. For example, it is now clear that data-driven and physics-based
predictive models are intimately related \cite{rupe22b}. Beyond predictive
modeling, uncovering the underlying mechanisms behind emergent behaviors is a
crucial component of scientific inquiry and remains an outstanding challenge.
Physics-informed data-driven methods, like those introduced here, provide a promising new avenue in this \cite{rung15a,Klus19a}.

Motivated by local causality and the physics of organization encapsulated by
causal equivalence, the local causal states provide universal representations for emergent organization in complex spatiotemporal systems. Our approximation algorithm and high-performance computing implementation make the local causal states actionable in practice on large-scale cutting-edge domain science problems. Here, we produced a quantitative verification of coherent structure detection by automatically counting vortices in free-decay two-dimensional turbulence and recovering the known-power law decay over time using the local causal states and their ability to capture nonstationary behavior. We then employed the local causal states for automated, fully-unsupervised extreme weather discovery and demonstrated their ability to extract both known and novel structures in global climate data. With further refinement, the local causal states may be able to additionally provide a scale for strength and intensity of identified extreme weather events \cite{ralph19a}. 

Further improvements to our extreme weather discovery may include the extension to other climate variable inputs, such as potential vorticity~\cite{wetz17a}. Similarly, additional vertical structure in the input variables could be included. For example, the IVT field is integrated over full columns, whereas our IVT-inspired multivariate analysis utilizes wind fields from mid-column only. As described in detail in Ref.~\cite{Rupe19a}, our current HPC implementation is heavily memory bound, limiting the amount of input data. While the current implementation can scale to a large amount of input data on a large HPC system \cite{Rupe19a}, scaling may be further improved in the future, for example with development of a streaming and distributed K-Means algorithm. 

Taken all together, the local causal states bridge the gap between theoretically-motivated organizational principles and domain science applications, where they perform comparably to or even better than specialized heuristics catering to the particular problem on hand.

\bibliography{chaos,spacetime}

\begin{thebibliography}{81}%
\makeatletter
\providecommand \@ifxundefined [1]{%
 \@ifx{#1\undefined}
}%
\providecommand \@ifnum [1]{%
 \ifnum #1\expandafter \@firstoftwo
 \else \expandafter \@secondoftwo
 \fi
}%
\providecommand \@ifx [1]{%
 \ifx #1\expandafter \@firstoftwo
 \else \expandafter \@secondoftwo
 \fi
}%
\providecommand \natexlab [1]{#1}%
\providecommand \enquote  [1]{``#1''}%
\providecommand \bibnamefont  [1]{#1}%
\providecommand \bibfnamefont [1]{#1}%
\providecommand \citenamefont [1]{#1}%
\providecommand \href@noop [0]{\@secondoftwo}%
\providecommand \href [0]{\begingroup \@sanitize@url \@href}%
\providecommand \@href[1]{\@@startlink{#1}\@@href}%
\providecommand \@@href[1]{\endgroup#1\@@endlink}%
\providecommand \@sanitize@url [0]{\catcode `\\12\catcode `\$12\catcode
  `\&12\catcode `\#12\catcode `\^12\catcode `\_12\catcode `\%12\relax}%
\providecommand \@@startlink[1]{}%
\providecommand \@@endlink[0]{}%
\providecommand \url  [0]{\begingroup\@sanitize@url \@url }%
\providecommand \@url [1]{\endgroup\@href {#1}{\urlprefix }}%
\providecommand \urlprefix  [0]{URL }%
\providecommand \Eprint [0]{\href }%
\providecommand \doibase [0]{https://doi.org/}%
\providecommand \selectlanguage [0]{\@gobble}%
\providecommand \bibinfo  [0]{\@secondoftwo}%
\providecommand \bibfield  [0]{\@secondoftwo}%
\providecommand \translation [1]{[#1]}%
\providecommand \BibitemOpen [0]{}%
\providecommand \bibitemStop [0]{}%
\providecommand \bibitemNoStop [0]{.\EOS\space}%
\providecommand \EOS [0]{\spacefactor3000\relax}%
\providecommand \BibitemShut  [1]{\csname bibitem#1\endcsname}%
\let\auto@bib@innerbib\@empty
\bibitem [{\citenamefont {Haller}(2015)}]{Hall15a}%
  \BibitemOpen
  \bibfield  {author} {\bibinfo {author} {\bibfnamefont {G.}~\bibnamefont
  {Haller}},\ }\bibfield  {title} {\bibinfo {title} {Lagrangian coherent
  structures},\ }\href@noop {} {\bibfield  {journal} {\bibinfo  {journal} {Ann.
  Rev. Fluid Mech.}\ }\textbf {\bibinfo {volume} {47}},\ \bibinfo {pages} {137}
  (\bibinfo {year} {2015})}\BibitemShut {NoStop}%
\bibitem [{\citenamefont {Holmes}\ \emph {et~al.}(2012)\citenamefont {Holmes},
  \citenamefont {Lumley}, \citenamefont {Berkooz},\ and\ \citenamefont
  {Rowley}}]{Holm12a}%
  \BibitemOpen
  \bibfield  {author} {\bibinfo {author} {\bibfnamefont {P.}~\bibnamefont
  {Holmes}}, \bibinfo {author} {\bibfnamefont {J.}~\bibnamefont {Lumley}},
  \bibinfo {author} {\bibfnamefont {G.}~\bibnamefont {Berkooz}},\ and\ \bibinfo
  {author} {\bibfnamefont {C.}~\bibnamefont {Rowley}},\ }\href@noop {} {\emph
  {\bibinfo {title} {Turbulence, Coherent Structures, Dynamical Systems and
  Symmetry}}}\ (\bibinfo  {publisher} {Cambridge University Press},\ \bibinfo
  {address} {Cambridge, United Kingdom},\ \bibinfo {year} {2012})\BibitemShut
  {NoStop}%
\bibitem [{\citenamefont {Mezi{\'c}}(2013)}]{Mezi13a}%
  \BibitemOpen
  \bibfield  {author} {\bibinfo {author} {\bibfnamefont {I.}~\bibnamefont
  {Mezi{\'c}}},\ }\bibfield  {title} {\bibinfo {title} {Analysis of fluid flows
  via spectral properties of the koopman operator},\ }\href@noop {} {\bibfield
  {journal} {\bibinfo  {journal} {Annual Review of Fluid Mechanics}\ }\textbf
  {\bibinfo {volume} {45}},\ \bibinfo {pages} {357} (\bibinfo {year}
  {2013})}\BibitemShut {NoStop}%
\bibitem [{\citenamefont {Hadjighasem}\ \emph {et~al.}(2017)\citenamefont
  {Hadjighasem}, \citenamefont {Farazmand}, \citenamefont {Blazevski},
  \citenamefont {Froyland},\ and\ \citenamefont {Haller}}]{Hadj17a}%
  \BibitemOpen
  \bibfield  {author} {\bibinfo {author} {\bibfnamefont {A.}~\bibnamefont
  {Hadjighasem}}, \bibinfo {author} {\bibfnamefont {M.}~\bibnamefont
  {Farazmand}}, \bibinfo {author} {\bibfnamefont {D.}~\bibnamefont
  {Blazevski}}, \bibinfo {author} {\bibfnamefont {G.}~\bibnamefont
  {Froyland}},\ and\ \bibinfo {author} {\bibfnamefont {G.}~\bibnamefont
  {Haller}},\ }\bibfield  {title} {\bibinfo {title} {A critical comparison of
  {L}agrangian methods for coherent structure detection},\ }\href@noop {}
  {\bibfield  {journal} {\bibinfo  {journal} {Chaos}\ }\textbf {\bibinfo
  {volume} {27}},\ \bibinfo {pages} {053104} (\bibinfo {year}
  {2017})}\BibitemShut {NoStop}%
\bibitem [{\citenamefont {Ghil}\ and\ \citenamefont
  {Lucarini}(2020)}]{ghil20a}%
  \BibitemOpen
  \bibfield  {author} {\bibinfo {author} {\bibfnamefont {M.}~\bibnamefont
  {Ghil}}\ and\ \bibinfo {author} {\bibfnamefont {V.}~\bibnamefont
  {Lucarini}},\ }\bibfield  {title} {\bibinfo {title} {The physics of climate
  variability and climate change},\ }\href@noop {} {\bibfield  {journal}
  {\bibinfo  {journal} {Rev. of Mod. Phys.}\ }\textbf {\bibinfo {volume}
  {92}},\ \bibinfo {pages} {035002} (\bibinfo {year} {2020})}\BibitemShut
  {NoStop}%
\bibitem [{\citenamefont {Ravishankara}\ \emph {et~al.}(2022)\citenamefont
  {Ravishankara}, \citenamefont {Randall},\ and\ \citenamefont
  {Hurrell}}]{Ravi22a}%
  \BibitemOpen
  \bibfield  {author} {\bibinfo {author} {\bibfnamefont {A.~R.}\ \bibnamefont
  {Ravishankara}}, \bibinfo {author} {\bibfnamefont {D.~A.}\ \bibnamefont
  {Randall}},\ and\ \bibinfo {author} {\bibfnamefont {J.~W.}\ \bibnamefont
  {Hurrell}},\ }\bibfield  {title} {\bibinfo {title} {Complex and yet
  predictable: The message of the 2021 {Nobel Prize} in physics},\ }\href
  {https://doi.org/10.1073/pnas.2120669119} {\bibfield  {journal} {\bibinfo
  {journal} {Proc. Natl. Acad. Sci. USA}\ }\textbf {\bibinfo {volume} {119}},\
  \bibinfo {pages} {e2120669119} (\bibinfo {year} {2022})}\BibitemShut
  {NoStop}%
\bibitem [{\citenamefont {P{\"o}rtner}\ \emph {et~al.}(2022)\citenamefont
  {P{\"o}rtner}, \citenamefont {Roberts}, \citenamefont {Adams}, \citenamefont
  {Adler}, \citenamefont {Aldunce}, \citenamefont {Ali}, \citenamefont {Begum},
  \citenamefont {Betts}, \citenamefont {Kerr}, \citenamefont {Biesbroek} \emph
  {et~al.}}]{ipcc22a}%
  \BibitemOpen
  \bibfield  {author} {\bibinfo {author} {\bibfnamefont {H.-O.}\ \bibnamefont
  {P{\"o}rtner}}, \bibinfo {author} {\bibfnamefont {D.~C.}\ \bibnamefont
  {Roberts}}, \bibinfo {author} {\bibfnamefont {H.}~\bibnamefont {Adams}},
  \bibinfo {author} {\bibfnamefont {C.}~\bibnamefont {Adler}}, \bibinfo
  {author} {\bibfnamefont {P.}~\bibnamefont {Aldunce}}, \bibinfo {author}
  {\bibfnamefont {E.}~\bibnamefont {Ali}}, \bibinfo {author} {\bibfnamefont
  {R.~A.}\ \bibnamefont {Begum}}, \bibinfo {author} {\bibfnamefont
  {R.}~\bibnamefont {Betts}}, \bibinfo {author} {\bibfnamefont {R.~B.}\
  \bibnamefont {Kerr}}, \bibinfo {author} {\bibfnamefont {R.}~\bibnamefont
  {Biesbroek}}, \emph {et~al.},\ }\href@noop {} {\emph {\bibinfo {title}
  {Climate change 2022: Impacts, adaptation and vulnerability}}}\ (\bibinfo
  {publisher} {{IPCC} {G}eneva, {S}witzerland:},\ \bibinfo {year}
  {2022})\BibitemShut {NoStop}%
\bibitem [{\citenamefont {Tippett}(2018)}]{tipp18a}%
  \BibitemOpen
  \bibfield  {author} {\bibinfo {author} {\bibfnamefont {M.~K.}\ \bibnamefont
  {Tippett}},\ }\bibfield  {title} {\bibinfo {title} {Extreme weather and
  climate},\ }\href@noop {} {\bibfield  {journal} {\bibinfo  {journal} {npj
  Climate and Atmospheric Science}\ }\textbf {\bibinfo {volume} {1}},\ \bibinfo
  {pages} {45} (\bibinfo {year} {2018})}\BibitemShut {NoStop}%
\bibitem [{\citenamefont {Roberts}\ \emph {et~al.}(2020)\citenamefont
  {Roberts}, \citenamefont {Camp}, \citenamefont {Seddon}, \citenamefont
  {Vidale}, \citenamefont {Hodges}, \citenamefont {Vannière}, \citenamefont
  {Mecking}, \citenamefont {Haarsma}, \citenamefont {Bellucci}, \citenamefont
  {Scoccimarro}, \citenamefont {Caron}, \citenamefont {Chauvin}, \citenamefont
  {Terray}, \citenamefont {Valcke}, \citenamefont {Moine}, \citenamefont
  {Putrasahan}, \citenamefont {Roberts}, \citenamefont {Senan}, \citenamefont
  {Zarzycki}, \citenamefont {Ullrich}, \citenamefont {Yamada}, \citenamefont
  {Mizuta}, \citenamefont {Kodama}, \citenamefont {Fu}, \citenamefont {Zhang},
  \citenamefont {Danabasoglu}, \citenamefont {Rosenbloom}, \citenamefont
  {Wang},\ and\ \citenamefont {Wu}}]{Rob20a}%
  \BibitemOpen
  \bibfield  {author} {\bibinfo {author} {\bibfnamefont {M.~J.}\ \bibnamefont
  {Roberts}}, \bibinfo {author} {\bibfnamefont {J.}~\bibnamefont {Camp}},
  \bibinfo {author} {\bibfnamefont {J.}~\bibnamefont {Seddon}}, \bibinfo
  {author} {\bibfnamefont {P.~L.}\ \bibnamefont {Vidale}}, \bibinfo {author}
  {\bibfnamefont {K.}~\bibnamefont {Hodges}}, \bibinfo {author} {\bibfnamefont
  {B.}~\bibnamefont {Vannière}}, \bibinfo {author} {\bibfnamefont
  {J.}~\bibnamefont {Mecking}}, \bibinfo {author} {\bibfnamefont
  {R.}~\bibnamefont {Haarsma}}, \bibinfo {author} {\bibfnamefont
  {A.}~\bibnamefont {Bellucci}}, \bibinfo {author} {\bibfnamefont
  {E.}~\bibnamefont {Scoccimarro}}, \bibinfo {author} {\bibfnamefont {L.-P.}\
  \bibnamefont {Caron}}, \bibinfo {author} {\bibfnamefont {F.}~\bibnamefont
  {Chauvin}}, \bibinfo {author} {\bibfnamefont {L.}~\bibnamefont {Terray}},
  \bibinfo {author} {\bibfnamefont {S.}~\bibnamefont {Valcke}}, \bibinfo
  {author} {\bibfnamefont {M.-P.}\ \bibnamefont {Moine}}, \bibinfo {author}
  {\bibfnamefont {D.}~\bibnamefont {Putrasahan}}, \bibinfo {author}
  {\bibfnamefont {C.~D.}\ \bibnamefont {Roberts}}, \bibinfo {author}
  {\bibfnamefont {R.}~\bibnamefont {Senan}}, \bibinfo {author} {\bibfnamefont
  {C.}~\bibnamefont {Zarzycki}}, \bibinfo {author} {\bibfnamefont
  {P.}~\bibnamefont {Ullrich}}, \bibinfo {author} {\bibfnamefont
  {Y.}~\bibnamefont {Yamada}}, \bibinfo {author} {\bibfnamefont
  {R.}~\bibnamefont {Mizuta}}, \bibinfo {author} {\bibfnamefont
  {C.}~\bibnamefont {Kodama}}, \bibinfo {author} {\bibfnamefont
  {D.}~\bibnamefont {Fu}}, \bibinfo {author} {\bibfnamefont {Q.}~\bibnamefont
  {Zhang}}, \bibinfo {author} {\bibfnamefont {G.}~\bibnamefont {Danabasoglu}},
  \bibinfo {author} {\bibfnamefont {N.}~\bibnamefont {Rosenbloom}}, \bibinfo
  {author} {\bibfnamefont {H.}~\bibnamefont {Wang}},\ and\ \bibinfo {author}
  {\bibfnamefont {L.}~\bibnamefont {Wu}},\ }\bibfield  {title} {\bibinfo
  {title} {Projected future changes in tropical cyclones using the {CMIP6
  HighResMIP} multimodel ensemble},\ }\href
  {https://doi.org/https://doi.org/10.1029/2020GL088662} {\bibfield  {journal}
  {\bibinfo  {journal} {Geophysical Research Letters}\ }\textbf {\bibinfo
  {volume} {47}},\ \bibinfo {pages} {e2020GL088662} (\bibinfo {year}
  {2020})}\BibitemShut {NoStop}%
\bibitem [{\citenamefont {Robinson}\ \emph {et~al.}(2021)\citenamefont
  {Robinson}, \citenamefont {Lehmann}, \citenamefont {Barriopedro},
  \citenamefont {Rahmstorf},\ and\ \citenamefont {Coumou}}]{robi21a}%
  \BibitemOpen
  \bibfield  {author} {\bibinfo {author} {\bibfnamefont {A.}~\bibnamefont
  {Robinson}}, \bibinfo {author} {\bibfnamefont {J.}~\bibnamefont {Lehmann}},
  \bibinfo {author} {\bibfnamefont {D.}~\bibnamefont {Barriopedro}}, \bibinfo
  {author} {\bibfnamefont {S.}~\bibnamefont {Rahmstorf}},\ and\ \bibinfo
  {author} {\bibfnamefont {D.}~\bibnamefont {Coumou}},\ }\bibfield  {title}
  {\bibinfo {title} {Increasing heat and rainfall extremes now far outside the
  historical climate},\ }\href@noop {} {\bibfield  {journal} {\bibinfo
  {journal} {npj Climate and Atmospheric Science}\ }\textbf {\bibinfo {volume}
  {4}},\ \bibinfo {pages} {45} (\bibinfo {year} {2021})}\BibitemShut {NoStop}%
\bibitem [{\citenamefont {Edwards}(2010)}]{edwa10a}%
  \BibitemOpen
  \bibfield  {author} {\bibinfo {author} {\bibfnamefont {P.~N.}\ \bibnamefont
  {Edwards}},\ }\href@noop {} {\emph {\bibinfo {title} {A Vast Machine:
  Computer Models, Climate Data, and the Politics of Global Warming}}}\
  (\bibinfo  {publisher} {The MIT Press},\ \bibinfo {year} {2010})\BibitemShut
  {NoStop}%
\bibitem [{\citenamefont {Flato}\ \emph {et~al.}(2013)\citenamefont {Flato},
  \citenamefont {Marotzke}, \citenamefont {Abiodun}, \citenamefont {Braconnot},
  \citenamefont {Chou}, \citenamefont {Collins}, \citenamefont {Cox},
  \citenamefont {Driouech}, \citenamefont {Emori}, \citenamefont {Eyring},
  \citenamefont {Forest}, \citenamefont {Gleckler}, \citenamefont {Guilyardi},
  \citenamefont {Jakob}, \citenamefont {Kattsov}, \citenamefont {Reason},\ and\
  \citenamefont {Rummukainen}}]{ipcc13a}%
  \BibitemOpen
  \bibfield  {author} {\bibinfo {author} {\bibfnamefont {G.}~\bibnamefont
  {Flato}}, \bibinfo {author} {\bibfnamefont {J.}~\bibnamefont {Marotzke}},
  \bibinfo {author} {\bibfnamefont {B.}~\bibnamefont {Abiodun}}, \bibinfo
  {author} {\bibfnamefont {P.}~\bibnamefont {Braconnot}}, \bibinfo {author}
  {\bibfnamefont {S.}~\bibnamefont {Chou}}, \bibinfo {author} {\bibfnamefont
  {W.}~\bibnamefont {Collins}}, \bibinfo {author} {\bibfnamefont
  {P.}~\bibnamefont {Cox}}, \bibinfo {author} {\bibfnamefont {F.}~\bibnamefont
  {Driouech}}, \bibinfo {author} {\bibfnamefont {S.}~\bibnamefont {Emori}},
  \bibinfo {author} {\bibfnamefont {V.}~\bibnamefont {Eyring}}, \bibinfo
  {author} {\bibfnamefont {C.}~\bibnamefont {Forest}}, \bibinfo {author}
  {\bibfnamefont {P.}~\bibnamefont {Gleckler}}, \bibinfo {author}
  {\bibfnamefont {E.}~\bibnamefont {Guilyardi}}, \bibinfo {author}
  {\bibfnamefont {C.}~\bibnamefont {Jakob}}, \bibinfo {author} {\bibfnamefont
  {V.}~\bibnamefont {Kattsov}}, \bibinfo {author} {\bibfnamefont
  {C.}~\bibnamefont {Reason}},\ and\ \bibinfo {author} {\bibfnamefont
  {M.}~\bibnamefont {Rummukainen}},\ }\bibinfo {title} {Evaluation of climate
  models},\ in\ \href@noop {} {\emph {\bibinfo {booktitle} {Climate Change
  2013: The Physical Science Basis. Contribution of Working Group I to the
  Fifth Assessment Report of the Intergovernmental Panel on Climate Change}}},\
  \bibinfo {editor} {edited by\ \bibinfo {editor} {\bibfnamefont
  {T.}~\bibnamefont {Stocker}}, \bibinfo {editor} {\bibfnamefont
  {D.}~\bibnamefont {Qin}}, \bibinfo {editor} {\bibfnamefont {G.-K.}\
  \bibnamefont {Plattner}}, \bibinfo {editor} {\bibfnamefont {M.}~\bibnamefont
  {Tignor}}, \bibinfo {editor} {\bibfnamefont {S.}~\bibnamefont {Allen}},
  \bibinfo {editor} {\bibfnamefont {J.}~\bibnamefont {Boschung}}, \bibinfo
  {editor} {\bibfnamefont {A.}~\bibnamefont {Nauels}}, \bibinfo {editor}
  {\bibfnamefont {Y.}~\bibnamefont {Xia}}, \bibinfo {editor} {\bibfnamefont
  {V.}~\bibnamefont {Bex}},\ and\ \bibinfo {editor} {\bibfnamefont
  {P.}~\bibnamefont {Midgley}}}\ (\bibinfo  {publisher} {Cambridge University
  Press},\ \bibinfo {address} {Cambridge, United Kingdom and New York, NY,
  USA},\ \bibinfo {year} {2013})\BibitemShut {NoStop}%
\bibitem [{\citenamefont {Eyring}\ \emph {et~al.}(2016)\citenamefont {Eyring},
  \citenamefont {Bony}, \citenamefont {Meehl}, \citenamefont {Senior},
  \citenamefont {Stevens}, \citenamefont {Stouffer},\ and\ \citenamefont
  {Taylor}}]{eyri16a}%
  \BibitemOpen
  \bibfield  {author} {\bibinfo {author} {\bibfnamefont {V.}~\bibnamefont
  {Eyring}}, \bibinfo {author} {\bibfnamefont {S.}~\bibnamefont {Bony}},
  \bibinfo {author} {\bibfnamefont {G.~A.}\ \bibnamefont {Meehl}}, \bibinfo
  {author} {\bibfnamefont {C.~A.}\ \bibnamefont {Senior}}, \bibinfo {author}
  {\bibfnamefont {B.}~\bibnamefont {Stevens}}, \bibinfo {author} {\bibfnamefont
  {R.~J.}\ \bibnamefont {Stouffer}},\ and\ \bibinfo {author} {\bibfnamefont
  {K.~E.}\ \bibnamefont {Taylor}},\ }\bibfield  {title} {\bibinfo {title}
  {Overview of the {C}oupled {M}odel {I}ntercomparison {P}roject {P}hase 6
  ({CMIP6}) experimental design and organization},\ }\href
  {https://doi.org/10.5194/gmd-9-1937-2016} {\bibfield  {journal} {\bibinfo
  {journal} {Geoscientific Model Development}\ }\textbf {\bibinfo {volume}
  {9}},\ \bibinfo {pages} {1937} (\bibinfo {year} {2016})}\BibitemShut
  {NoStop}%
\bibitem [{\citenamefont {Stockhause}\ and\ \citenamefont
  {Lautenschlager}(2017)}]{stoc17a}%
  \BibitemOpen
  \bibfield  {author} {\bibinfo {author} {\bibfnamefont {M.}~\bibnamefont
  {Stockhause}}\ and\ \bibinfo {author} {\bibfnamefont {M.}~\bibnamefont
  {Lautenschlager}},\ }\bibfield  {title} {\bibinfo {title} {{CMIP6} data
  citation of evolving data},\ }\href@noop {} {\bibfield  {journal} {\bibinfo
  {journal} {Data Science Journal}\ }\textbf {\bibinfo {volume} {16}} (\bibinfo
  {year} {2017})}\BibitemShut {NoStop}%
\bibitem [{\citenamefont {Neu}\ \emph {et~al.}(2013)\citenamefont {Neu},
  \citenamefont {Akperov}, \citenamefont {Bellenbaum}, \citenamefont
  {Benestad}, \citenamefont {Blender}, \citenamefont {Caballero}, \citenamefont
  {Cocozza}, \citenamefont {Dacre}, \citenamefont {Feng}, \citenamefont
  {Fraedrich}, \citenamefont {Grieger}, \citenamefont {Gulev}, \citenamefont
  {Hanley}, \citenamefont {Hewson}, \citenamefont {Inatsu}, \citenamefont
  {Keay}, \citenamefont {Kew}, \citenamefont {Kindem}, \citenamefont
  {Leckebusch}, \citenamefont {Liberato}, \citenamefont {Lionello},
  \citenamefont {Mokhov}, \citenamefont {Pinto}, \citenamefont {Raible},
  \citenamefont {Reale}, \citenamefont {Rudeva}, \citenamefont {Schuster},
  \citenamefont {Simmonds}, \citenamefont {Sinclair}, \citenamefont {Sprenger},
  \citenamefont {Tilinina}, \citenamefont {Trigo}, \citenamefont {Ulbrich},
  \citenamefont {Ulbrich}, \citenamefont {Wang},\ and\ \citenamefont
  {Wernli}}]{Neu13a}%
  \BibitemOpen
  \bibfield  {author} {\bibinfo {author} {\bibfnamefont {U.}~\bibnamefont
  {Neu}}, \bibinfo {author} {\bibfnamefont {M.~G.}\ \bibnamefont {Akperov}},
  \bibinfo {author} {\bibfnamefont {N.}~\bibnamefont {Bellenbaum}}, \bibinfo
  {author} {\bibfnamefont {R.}~\bibnamefont {Benestad}}, \bibinfo {author}
  {\bibfnamefont {R.}~\bibnamefont {Blender}}, \bibinfo {author} {\bibfnamefont
  {R.}~\bibnamefont {Caballero}}, \bibinfo {author} {\bibfnamefont
  {A.}~\bibnamefont {Cocozza}}, \bibinfo {author} {\bibfnamefont {H.~F.}\
  \bibnamefont {Dacre}}, \bibinfo {author} {\bibfnamefont {Y.}~\bibnamefont
  {Feng}}, \bibinfo {author} {\bibfnamefont {K.}~\bibnamefont {Fraedrich}},
  \bibinfo {author} {\bibfnamefont {J.}~\bibnamefont {Grieger}}, \bibinfo
  {author} {\bibfnamefont {S.}~\bibnamefont {Gulev}}, \bibinfo {author}
  {\bibfnamefont {J.}~\bibnamefont {Hanley}}, \bibinfo {author} {\bibfnamefont
  {T.}~\bibnamefont {Hewson}}, \bibinfo {author} {\bibfnamefont
  {M.}~\bibnamefont {Inatsu}}, \bibinfo {author} {\bibfnamefont
  {K.}~\bibnamefont {Keay}}, \bibinfo {author} {\bibfnamefont {S.~F.}\
  \bibnamefont {Kew}}, \bibinfo {author} {\bibfnamefont {I.}~\bibnamefont
  {Kindem}}, \bibinfo {author} {\bibfnamefont {G.~C.}\ \bibnamefont
  {Leckebusch}}, \bibinfo {author} {\bibfnamefont {M.~L.~R.}\ \bibnamefont
  {Liberato}}, \bibinfo {author} {\bibfnamefont {P.}~\bibnamefont {Lionello}},
  \bibinfo {author} {\bibfnamefont {I.~I.}\ \bibnamefont {Mokhov}}, \bibinfo
  {author} {\bibfnamefont {J.~G.}\ \bibnamefont {Pinto}}, \bibinfo {author}
  {\bibfnamefont {C.~C.}\ \bibnamefont {Raible}}, \bibinfo {author}
  {\bibfnamefont {M.}~\bibnamefont {Reale}}, \bibinfo {author} {\bibfnamefont
  {I.}~\bibnamefont {Rudeva}}, \bibinfo {author} {\bibfnamefont
  {M.}~\bibnamefont {Schuster}}, \bibinfo {author} {\bibfnamefont
  {I.}~\bibnamefont {Simmonds}}, \bibinfo {author} {\bibfnamefont
  {M.}~\bibnamefont {Sinclair}}, \bibinfo {author} {\bibfnamefont
  {M.}~\bibnamefont {Sprenger}}, \bibinfo {author} {\bibfnamefont {N.~D.}\
  \bibnamefont {Tilinina}}, \bibinfo {author} {\bibfnamefont {I.~F.}\
  \bibnamefont {Trigo}}, \bibinfo {author} {\bibfnamefont {S.}~\bibnamefont
  {Ulbrich}}, \bibinfo {author} {\bibfnamefont {U.}~\bibnamefont {Ulbrich}},
  \bibinfo {author} {\bibfnamefont {X.~L.}\ \bibnamefont {Wang}},\ and\
  \bibinfo {author} {\bibfnamefont {H.}~\bibnamefont {Wernli}},\ }\bibfield
  {title} {\bibinfo {title} {{IMILAST}: A community effort to intercompare
  extratropical cyclone detection and tracking algorithms:},\ }\href
  {https://doi.org/https://doi.org/10.1175/BAMS-D-11-00154.1} {\bibfield
  {journal} {\bibinfo  {journal} {Bulletin of the American Meteorological
  Society}\ }\textbf {\bibinfo {volume} {94}},\ \bibinfo {pages} {529 }
  (\bibinfo {year} {2013})}\BibitemShut {NoStop}%
\bibitem [{\citenamefont {Shields}\ \emph {et~al.}(2018)\citenamefont
  {Shields}, \citenamefont {Rutz}, \citenamefont {Leung}, \citenamefont
  {Ralph}, \citenamefont {Wehner}, \citenamefont {Kawzenuk}, \citenamefont
  {Lora}, \citenamefont {McClenny}, \citenamefont {Osborne}, \citenamefont
  {Payne}, \citenamefont {Ullrich}, \citenamefont {Gershunov}, \citenamefont
  {Goldenson}, \citenamefont {Guan}, \citenamefont {Qian}, \citenamefont
  {Ramos}, \citenamefont {Sarangi}, \citenamefont {Sellars}, \citenamefont
  {Gorodetskaya}, \citenamefont {Kashinath}, \citenamefont {Kurlin},
  \citenamefont {Mahoney}, \citenamefont {Muszynski}, \citenamefont {Pierce},
  \citenamefont {Subramanian}, \citenamefont {Tome}, \citenamefont {Waliser},
  \citenamefont {Walton}, \citenamefont {Wick}, \citenamefont {Wilson},
  \citenamefont {Lavers}, \citenamefont {Prabhat}, \citenamefont {Collow},
  \citenamefont {Krishnan}, \citenamefont {Magnusdottir},\ and\ \citenamefont
  {Nguyen}}]{Shie18a}%
  \BibitemOpen
  \bibfield  {author} {\bibinfo {author} {\bibfnamefont {C.~A.}\ \bibnamefont
  {Shields}}, \bibinfo {author} {\bibfnamefont {J.~J.}\ \bibnamefont {Rutz}},
  \bibinfo {author} {\bibfnamefont {L.-Y.}\ \bibnamefont {Leung}}, \bibinfo
  {author} {\bibfnamefont {F.~M.}\ \bibnamefont {Ralph}}, \bibinfo {author}
  {\bibfnamefont {M.}~\bibnamefont {Wehner}}, \bibinfo {author} {\bibfnamefont
  {B.}~\bibnamefont {Kawzenuk}}, \bibinfo {author} {\bibfnamefont {J.~M.}\
  \bibnamefont {Lora}}, \bibinfo {author} {\bibfnamefont {E.}~\bibnamefont
  {McClenny}}, \bibinfo {author} {\bibfnamefont {T.}~\bibnamefont {Osborne}},
  \bibinfo {author} {\bibfnamefont {A.~E.}\ \bibnamefont {Payne}}, \bibinfo
  {author} {\bibfnamefont {P.}~\bibnamefont {Ullrich}}, \bibinfo {author}
  {\bibfnamefont {A.}~\bibnamefont {Gershunov}}, \bibinfo {author}
  {\bibfnamefont {N.}~\bibnamefont {Goldenson}}, \bibinfo {author}
  {\bibfnamefont {B.}~\bibnamefont {Guan}}, \bibinfo {author} {\bibfnamefont
  {Y.}~\bibnamefont {Qian}}, \bibinfo {author} {\bibfnamefont {A.~M.}\
  \bibnamefont {Ramos}}, \bibinfo {author} {\bibfnamefont {C.}~\bibnamefont
  {Sarangi}}, \bibinfo {author} {\bibfnamefont {S.}~\bibnamefont {Sellars}},
  \bibinfo {author} {\bibfnamefont {I.}~\bibnamefont {Gorodetskaya}}, \bibinfo
  {author} {\bibfnamefont {K.}~\bibnamefont {Kashinath}}, \bibinfo {author}
  {\bibfnamefont {V.}~\bibnamefont {Kurlin}}, \bibinfo {author} {\bibfnamefont
  {K.}~\bibnamefont {Mahoney}}, \bibinfo {author} {\bibfnamefont
  {G.}~\bibnamefont {Muszynski}}, \bibinfo {author} {\bibfnamefont
  {R.}~\bibnamefont {Pierce}}, \bibinfo {author} {\bibfnamefont {A.~C.}\
  \bibnamefont {Subramanian}}, \bibinfo {author} {\bibfnamefont
  {R.}~\bibnamefont {Tome}}, \bibinfo {author} {\bibfnamefont {D.}~\bibnamefont
  {Waliser}}, \bibinfo {author} {\bibfnamefont {D.}~\bibnamefont {Walton}},
  \bibinfo {author} {\bibfnamefont {G.}~\bibnamefont {Wick}}, \bibinfo {author}
  {\bibfnamefont {A.}~\bibnamefont {Wilson}}, \bibinfo {author} {\bibfnamefont
  {D.}~\bibnamefont {Lavers}}, \bibinfo {author} {\bibnamefont {Prabhat}},
  \bibinfo {author} {\bibfnamefont {A.}~\bibnamefont {Collow}}, \bibinfo
  {author} {\bibfnamefont {H.}~\bibnamefont {Krishnan}}, \bibinfo {author}
  {\bibfnamefont {G.}~\bibnamefont {Magnusdottir}},\ and\ \bibinfo {author}
  {\bibfnamefont {P.}~\bibnamefont {Nguyen}},\ }\bibfield  {title} {\bibinfo
  {title} {Atmospheric river tracking method intercomparison project
  ({ARTMIP}): project goals and experimental design},\ }\href
  {https://doi.org/10.5194/gmd-11-2455-2018} {\bibfield  {journal} {\bibinfo
  {journal} {Geoscientific Model Development}\ }\textbf {\bibinfo {volume}
  {11}},\ \bibinfo {pages} {2455} (\bibinfo {year} {2018})}\BibitemShut
  {NoStop}%
\bibitem [{\citenamefont {Mudigonda}\ \emph {et~al.}(2017)\citenamefont
  {Mudigonda}, \citenamefont {Kim}, \citenamefont {Mahesh}, \citenamefont
  {.Kahou}, \citenamefont {Kashinath}, \citenamefont {Williams}, \citenamefont
  {Michalski}, \citenamefont {O’Brien},\ and\ \citenamefont
  {Prabhat}}]{mudi17a}%
  \BibitemOpen
  \bibfield  {author} {\bibinfo {author} {\bibfnamefont {M.}~\bibnamefont
  {Mudigonda}}, \bibinfo {author} {\bibfnamefont {S.}~\bibnamefont {Kim}},
  \bibinfo {author} {\bibfnamefont {A.}~\bibnamefont {Mahesh}}, \bibinfo
  {author} {\bibfnamefont {S.}~\bibnamefont {.Kahou}}, \bibinfo {author}
  {\bibfnamefont {K.}~\bibnamefont {Kashinath}}, \bibinfo {author}
  {\bibfnamefont {D.}~\bibnamefont {Williams}}, \bibinfo {author}
  {\bibfnamefont {V.}~\bibnamefont {Michalski}}, \bibinfo {author}
  {\bibfnamefont {T.}~\bibnamefont {O’Brien}},\ and\ \bibinfo {author}
  {\bibnamefont {Prabhat}},\ }\bibfield  {title} {\bibinfo {title} {Segmenting
  and tracking extreme climate events using neural networks},\ }in\ \href@noop
  {} {\emph {\bibinfo {booktitle} {DLPS Workshop, NeurIPS}}}\ (\bibinfo {year}
  {2017})\BibitemShut {NoStop}%
\bibitem [{\citenamefont {Jiang}\ \emph {et~al.}(2019)\citenamefont {Jiang},
  \citenamefont {Huang}, \citenamefont {Kashinath}, \citenamefont {Prabhat},
  \citenamefont {Marcus},\ and\ \citenamefont {Niessner}}]{jian18a}%
  \BibitemOpen
  \bibfield  {author} {\bibinfo {author} {\bibfnamefont {C.}~\bibnamefont
  {Jiang}}, \bibinfo {author} {\bibfnamefont {J.}~\bibnamefont {Huang}},
  \bibinfo {author} {\bibfnamefont {K.}~\bibnamefont {Kashinath}}, \bibinfo
  {author} {\bibnamefont {Prabhat}}, \bibinfo {author} {\bibfnamefont
  {P.}~\bibnamefont {Marcus}},\ and\ \bibinfo {author} {\bibfnamefont
  {M.}~\bibnamefont {Niessner}},\ }\bibfield  {title} {\bibinfo {title}
  {Spherical {CNN}s on unstructured grids},\ }in\ \href@noop {} {\emph
  {\bibinfo {booktitle} {International Conference on Learning
  Representations}}}\ (\bibinfo {year} {2019})\BibitemShut {NoStop}%
\bibitem [{\citenamefont {Cohen}\ \emph {et~al.}(2019)\citenamefont {Cohen},
  \citenamefont {Weiler}, \citenamefont {Kicanaoglu},\ and\ \citenamefont
  {Welling}}]{cohe19a}%
  \BibitemOpen
  \bibfield  {author} {\bibinfo {author} {\bibfnamefont {T.}~\bibnamefont
  {Cohen}}, \bibinfo {author} {\bibfnamefont {M.}~\bibnamefont {Weiler}},
  \bibinfo {author} {\bibfnamefont {B.}~\bibnamefont {Kicanaoglu}},\ and\
  \bibinfo {author} {\bibfnamefont {M.}~\bibnamefont {Welling}},\ }\bibfield
  {title} {\bibinfo {title} {Gauge equivariant convolutional networks and the
  icosahedral {CNN}},\ }in\ \href@noop {} {\emph {\bibinfo {booktitle}
  {Proceedings of the 36th International Conference on Machine Learning}}},\
  \bibinfo {series} {Proceedings of Machine Learning Research}, Vol.~\bibinfo
  {volume} {97},\ \bibinfo {editor} {edited by\ \bibinfo {editor}
  {\bibfnamefont {K.}~\bibnamefont {Chaudhuri}}\ and\ \bibinfo {editor}
  {\bibfnamefont {R.}~\bibnamefont {Salakhutdinov}}}\ (\bibinfo  {publisher}
  {PMLR},\ \bibinfo {address} {Long Beach, California, USA},\ \bibinfo {year}
  {2019})\ pp.\ \bibinfo {pages} {1321--1330}\BibitemShut {NoStop}%
\bibitem [{\citenamefont {Kurth}\ \emph {et~al.}(2018)\citenamefont {Kurth},
  \citenamefont {Treichler}, \citenamefont {Romero}, \citenamefont {Mudigonda},
  \citenamefont {Luehr}, \citenamefont {Phillips}, \citenamefont {Mahesh},
  \citenamefont {Matheson}, \citenamefont {Deslippe}, \citenamefont {Fatica},
  \citenamefont {Prabhat},\ and\ \citenamefont {Houston}}]{kurt18a}%
  \BibitemOpen
  \bibfield  {author} {\bibinfo {author} {\bibfnamefont {T.}~\bibnamefont
  {Kurth}}, \bibinfo {author} {\bibfnamefont {S.}~\bibnamefont {Treichler}},
  \bibinfo {author} {\bibfnamefont {J.}~\bibnamefont {Romero}}, \bibinfo
  {author} {\bibfnamefont {M.}~\bibnamefont {Mudigonda}}, \bibinfo {author}
  {\bibfnamefont {N.}~\bibnamefont {Luehr}}, \bibinfo {author} {\bibfnamefont
  {E.}~\bibnamefont {Phillips}}, \bibinfo {author} {\bibfnamefont
  {A.}~\bibnamefont {Mahesh}}, \bibinfo {author} {\bibfnamefont
  {M.}~\bibnamefont {Matheson}}, \bibinfo {author} {\bibfnamefont
  {J.}~\bibnamefont {Deslippe}}, \bibinfo {author} {\bibfnamefont
  {M.}~\bibnamefont {Fatica}}, \bibinfo {author} {\bibnamefont {Prabhat}},\
  and\ \bibinfo {author} {\bibfnamefont {M.}~\bibnamefont {Houston}},\
  }\bibfield  {title} {\bibinfo {title} {Exascale deep learning for climate
  analytics},\ }in\ \href@noop {} {\emph {\bibinfo {booktitle} {Proceedings of
  the International Conference for High Performance Computing, Networking,
  Storage, and Analysis}}}\ (\bibinfo {organization} {IEEE Press},\ \bibinfo
  {year} {2018})\ p.~\bibinfo {pages} {51}\BibitemShut {NoStop}%
\bibitem [{\citenamefont {Epps}(2017)}]{Epps17a}%
  \BibitemOpen
  \bibfield  {author} {\bibinfo {author} {\bibfnamefont {B.}~\bibnamefont
  {Epps}},\ }\bibfield  {title} {\bibinfo {title} {Review of vortex
  identification methods},\ }in\ \href@noop {} {\emph {\bibinfo {booktitle}
  {55th AIAA Aerospace Sciences Meeting}}}\ (\bibinfo {year} {2017})\ p.\
  \bibinfo {pages} {0989}\BibitemShut {NoStop}%
\bibitem [{\citenamefont {Prabhat}\ \emph {et~al.}(2012)\citenamefont
  {Prabhat}, \citenamefont {Rübel}, \citenamefont {Byna}, \citenamefont {Wu},
  \citenamefont {Li}, \citenamefont {Wehner}, \citenamefont {Bethel} \emph
  {et~al.}}]{Prab12a}%
  \BibitemOpen
  \bibfield  {author} {\bibinfo {author} {\bibnamefont {Prabhat}}, \bibinfo
  {author} {\bibfnamefont {O.}~\bibnamefont {Rübel}}, \bibinfo {author}
  {\bibfnamefont {S.}~\bibnamefont {Byna}}, \bibinfo {author} {\bibfnamefont
  {K.}~\bibnamefont {Wu}}, \bibinfo {author} {\bibfnamefont {F.}~\bibnamefont
  {Li}}, \bibinfo {author} {\bibfnamefont {M.}~\bibnamefont {Wehner}}, \bibinfo
  {author} {\bibfnamefont {W.}~\bibnamefont {Bethel}}, \emph {et~al.},\
  }\bibfield  {title} {\bibinfo {title} {{TECA}: A parallel toolkit for extreme
  climate analysis},\ }in\ \href@noop {} {\emph {\bibinfo {booktitle} {Third
  Worskhop on Data Mining in Earth System Science (DMESS)}}}\ (\bibinfo {year}
  {2012})\BibitemShut {NoStop}%
\bibitem [{\citenamefont {Prabhat}\ \emph {et~al.}(2021)\citenamefont
  {Prabhat}, \citenamefont {Kashinath}, \citenamefont {Mudigonda},
  \citenamefont {Kim}, \citenamefont {Kapp-Schwoerer}, \citenamefont
  {Graubner}, \citenamefont {Karaismailoglu}, \citenamefont {Von~Kleist},
  \citenamefont {Kurth}, \citenamefont {Greiner}, \citenamefont {Mahesh} \emph
  {et~al.}}]{prab21a}%
  \BibitemOpen
  \bibfield  {author} {\bibinfo {author} {\bibnamefont {Prabhat}}, \bibinfo
  {author} {\bibfnamefont {K.}~\bibnamefont {Kashinath}}, \bibinfo {author}
  {\bibfnamefont {M.}~\bibnamefont {Mudigonda}}, \bibinfo {author}
  {\bibfnamefont {S.}~\bibnamefont {Kim}}, \bibinfo {author} {\bibfnamefont
  {L.}~\bibnamefont {Kapp-Schwoerer}}, \bibinfo {author} {\bibfnamefont
  {A.}~\bibnamefont {Graubner}}, \bibinfo {author} {\bibfnamefont
  {E.}~\bibnamefont {Karaismailoglu}}, \bibinfo {author} {\bibfnamefont
  {L.}~\bibnamefont {Von~Kleist}}, \bibinfo {author} {\bibfnamefont
  {T.}~\bibnamefont {Kurth}}, \bibinfo {author} {\bibfnamefont
  {A.}~\bibnamefont {Greiner}}, \bibinfo {author} {\bibfnamefont
  {A.}~\bibnamefont {Mahesh}}, \emph {et~al.},\ }\bibfield  {title} {\bibinfo
  {title} {Climate{N}et: an expert-labeled open dataset and deep learning
  architecture for enabling high-precision analyses of extreme weather},\
  }\href@noop {} {\bibfield  {journal} {\bibinfo  {journal} {Geoscientific
  Model Development}\ }\textbf {\bibinfo {volume} {14}},\ \bibinfo {pages}
  {107} (\bibinfo {year} {2021})}\BibitemShut {NoStop}%
\bibitem [{\citenamefont {Crutchfield}(2012)}]{Crut12a}%
  \BibitemOpen
  \bibfield  {author} {\bibinfo {author} {\bibfnamefont {J.~P.}\ \bibnamefont
  {Crutchfield}},\ }\bibfield  {title} {\bibinfo {title} {Between order and
  chaos},\ }\href {https://doi.org/10.1038/NPHYS2190} {\bibfield  {journal}
  {\bibinfo  {journal} {Nature Physics}\ }\textbf {\bibinfo {volume} {8}},\
  \bibinfo {pages} {17} (\bibinfo {year} {2012})}\BibitemShut {NoStop}%
\bibitem [{\citenamefont {Rupe}\ and\ \citenamefont
  {Crutchfield}(2022)}]{rupe22a}%
  \BibitemOpen
  \bibfield  {author} {\bibinfo {author} {\bibfnamefont {A.}~\bibnamefont
  {Rupe}}\ and\ \bibinfo {author} {\bibfnamefont {J.~P.}\ \bibnamefont
  {Crutchfield}},\ }\bibfield  {title} {\bibinfo {title} {Algebraic theory of
  patterns as generalized symmetries},\ }\href@noop {} {\bibfield  {journal}
  {\bibinfo  {journal} {Symmetry}\ }\textbf {\bibinfo {volume} {14}},\ \bibinfo
  {pages} {1636} (\bibinfo {year} {2022})}\BibitemShut {NoStop}%
\bibitem [{\citenamefont {Rupe}\ and\ \citenamefont
  {Crutchfield}(2018)}]{Rupe18a}%
  \BibitemOpen
  \bibfield  {author} {\bibinfo {author} {\bibfnamefont {A.}~\bibnamefont
  {Rupe}}\ and\ \bibinfo {author} {\bibfnamefont {J.~P.}\ \bibnamefont
  {Crutchfield}},\ }\bibfield  {title} {\bibinfo {title} {Local causal states
  and discrete coherent structures},\ }\href
  {https://doi.org/10.1063/1.5021130} {\bibfield  {journal} {\bibinfo
  {journal} {Chaos}\ }\textbf {\bibinfo {volume} {28}},\ \bibinfo {pages} {1}
  (\bibinfo {year} {2018})}\BibitemShut {NoStop}%
\bibitem [{\citenamefont {Rupe}\ \emph {et~al.}(2019)\citenamefont {Rupe},
  \citenamefont {Kumar}, \citenamefont {Epifanov}, \citenamefont {Kashinath},
  \citenamefont {Pavlyk}, \citenamefont {Schlimbach}, \citenamefont {Patwary},
  \citenamefont {Maidanov}, \citenamefont {Lee}, \citenamefont {Prabhat},\ and\
  \citenamefont {Crutchfield}}]{Rupe19a}%
  \BibitemOpen
  \bibfield  {author} {\bibinfo {author} {\bibfnamefont {A.}~\bibnamefont
  {Rupe}}, \bibinfo {author} {\bibfnamefont {N.}~\bibnamefont {Kumar}},
  \bibinfo {author} {\bibfnamefont {V.}~\bibnamefont {Epifanov}}, \bibinfo
  {author} {\bibfnamefont {K.}~\bibnamefont {Kashinath}}, \bibinfo {author}
  {\bibfnamefont {O.}~\bibnamefont {Pavlyk}}, \bibinfo {author} {\bibfnamefont
  {F.}~\bibnamefont {Schlimbach}}, \bibinfo {author} {\bibfnamefont
  {M.}~\bibnamefont {Patwary}}, \bibinfo {author} {\bibfnamefont
  {S.}~\bibnamefont {Maidanov}}, \bibinfo {author} {\bibfnamefont
  {V.}~\bibnamefont {Lee}}, \bibinfo {author} {\bibnamefont {Prabhat}},\ and\
  \bibinfo {author} {\bibfnamefont {J.~P.}\ \bibnamefont {Crutchfield}},\
  }\bibfield  {title} {\bibinfo {title} {Disco: Physics-based unsupervised
  discovery of coherent structures in spatiotemporal systems},\ }in\ \href@noop
  {} {\emph {\bibinfo {booktitle} {2019 IEEE/ACM Workshop on Machine Learning
  in High Performance Computing Environments (MLHPC)}}}\ (\bibinfo
  {organization} {IEEE},\ \bibinfo {year} {2019})\ pp.\ \bibinfo {pages}
  {75--87}\BibitemShut {NoStop}%
\bibitem [{\citenamefont {Arthur}\ and\ \citenamefont
  {Vassilvitskii}(2007)}]{kmeans}%
  \BibitemOpen
  \bibfield  {author} {\bibinfo {author} {\bibfnamefont {D.}~\bibnamefont
  {Arthur}}\ and\ \bibinfo {author} {\bibfnamefont {S.}~\bibnamefont
  {Vassilvitskii}},\ }\bibfield  {title} {\bibinfo {title} {{K-Means++:} the
  advantages of careful seeding},\ }in\ \href@noop {} {\emph {\bibinfo
  {booktitle} {Proceedings of the eighteenth annual ACM-SIAM symposium on
  Discrete algorithms}}}\ (\bibinfo {organization} {Society for Industrial and
  Applied Mathematics},\ \bibinfo {year} {2007})\ pp.\ \bibinfo {pages}
  {1027--1035}\BibitemShut {NoStop}%
\bibitem [{\citenamefont {Froyland}\ \emph {et~al.}(2007)\citenamefont
  {Froyland}, \citenamefont {Padberg}, \citenamefont {England},\ and\
  \citenamefont {Treguier}}]{froy07a}%
  \BibitemOpen
  \bibfield  {author} {\bibinfo {author} {\bibfnamefont {G.}~\bibnamefont
  {Froyland}}, \bibinfo {author} {\bibfnamefont {K.}~\bibnamefont {Padberg}},
  \bibinfo {author} {\bibfnamefont {M.~H.}\ \bibnamefont {England}},\ and\
  \bibinfo {author} {\bibfnamefont {A.~M.}\ \bibnamefont {Treguier}},\
  }\bibfield  {title} {\bibinfo {title} {Detection of coherent oceanic
  structures via transfer operators},\ }\href@noop {} {\bibfield  {journal}
  {\bibinfo  {journal} {Physical review letters}\ }\textbf {\bibinfo {volume}
  {98}},\ \bibinfo {pages} {224503} (\bibinfo {year} {2007})}\BibitemShut
  {NoStop}%
\bibitem [{\citenamefont {Froyland}\ \emph {et~al.}(2010)\citenamefont
  {Froyland}, \citenamefont {Santitissadeekorn},\ and\ \citenamefont
  {Monahan}}]{froy10a}%
  \BibitemOpen
  \bibfield  {author} {\bibinfo {author} {\bibfnamefont {G.}~\bibnamefont
  {Froyland}}, \bibinfo {author} {\bibfnamefont {N.}~\bibnamefont
  {Santitissadeekorn}},\ and\ \bibinfo {author} {\bibfnamefont
  {A.}~\bibnamefont {Monahan}},\ }\bibfield  {title} {\bibinfo {title}
  {Transport in time-dependent dynamical systems: Finite-time coherent sets},\
  }\href@noop {} {\bibfield  {journal} {\bibinfo  {journal} {Chaos: An
  Interdisciplinary Journal of Nonlinear Science}\ }\textbf {\bibinfo {volume}
  {20}},\ \bibinfo {pages} {043116} (\bibinfo {year} {2010})}\BibitemShut
  {NoStop}%
\bibitem [{\citenamefont {McWilliams}(1990)}]{mcwil90a}%
  \BibitemOpen
  \bibfield  {author} {\bibinfo {author} {\bibfnamefont {J.~C.}\ \bibnamefont
  {McWilliams}},\ }\bibfield  {title} {\bibinfo {title} {The vortices of
  two-dimensional turbulence},\ }\href@noop {} {\bibfield  {journal} {\bibinfo
  {journal} {J. of Fluid Mech.}\ }\textbf {\bibinfo {volume} {219}},\ \bibinfo
  {pages} {361} (\bibinfo {year} {1990})}\BibitemShut {NoStop}%
\bibitem [{\citenamefont {Carnevale}\ \emph {et~al.}(1991)\citenamefont
  {Carnevale}, \citenamefont {McWilliams}, \citenamefont {Pomeau},
  \citenamefont {Weiss},\ and\ \citenamefont {Young}}]{carn91a}%
  \BibitemOpen
  \bibfield  {author} {\bibinfo {author} {\bibfnamefont {G.~F.}\ \bibnamefont
  {Carnevale}}, \bibinfo {author} {\bibfnamefont {J.~C.}\ \bibnamefont
  {McWilliams}}, \bibinfo {author} {\bibfnamefont {Y.}~\bibnamefont {Pomeau}},
  \bibinfo {author} {\bibfnamefont {J.~B.}\ \bibnamefont {Weiss}},\ and\
  \bibinfo {author} {\bibfnamefont {W.~R.}\ \bibnamefont {Young}},\ }\bibfield
  {title} {\bibinfo {title} {Evolution of vortex statistics in two-dimensional
  turbulence},\ }\href@noop {} {\bibfield  {journal} {\bibinfo  {journal}
  {Phys. Rev. Let.}\ }\textbf {\bibinfo {volume} {66}},\ \bibinfo {pages}
  {2735} (\bibinfo {year} {1991})}\BibitemShut {NoStop}%
\bibitem [{\citenamefont {Fiorio}\ and\ \citenamefont
  {Gustedt}(1996)}]{fior96a}%
  \BibitemOpen
  \bibfield  {author} {\bibinfo {author} {\bibfnamefont {C.}~\bibnamefont
  {Fiorio}}\ and\ \bibinfo {author} {\bibfnamefont {J.}~\bibnamefont
  {Gustedt}},\ }\bibfield  {title} {\bibinfo {title} {Two linear time
  union-find strategies for image processing},\ }\href@noop {} {\bibfield
  {journal} {\bibinfo  {journal} {Theoretical Computer Science}\ }\textbf
  {\bibinfo {volume} {154}},\ \bibinfo {pages} {165} (\bibinfo {year}
  {1996})}\BibitemShut {NoStop}%
\bibitem [{\citenamefont {Rupe}(2022{\natexlab{a}})}]{turbseg}%
  \BibitemOpen
  \bibfield  {author} {\bibinfo {author} {\bibfnamefont {A.}~\bibnamefont
  {Rupe}},\ }\href@noop {} {\bibinfo {title} {{2D} turbulence segmentation
  video}},\ \bibinfo {howpublished}
  {\url{https://drive.google.com/file/d/1RoAh6J_gadhDcgWFPXa5z6g9q55TDM7G/view}}
  (\bibinfo {year} {2022}{\natexlab{a}})\BibitemShut {NoStop}%
\bibitem [{\citenamefont {Froyland}\ and\ \citenamefont
  {Koltai}(2021)}]{froy21a}%
  \BibitemOpen
  \bibfield  {author} {\bibinfo {author} {\bibfnamefont {G.}~\bibnamefont
  {Froyland}}\ and\ \bibinfo {author} {\bibfnamefont {P.}~\bibnamefont
  {Koltai}},\ }\bibfield  {title} {\bibinfo {title} {Detecting the birth and
  death of finite-time coherent sets},\ }\href@noop {} {\bibfield  {journal}
  {\bibinfo  {journal} {arXiv preprint arXiv:2103.16286}\ } (\bibinfo {year}
  {2021})}\BibitemShut {NoStop}%
\bibitem [{\citenamefont {Wehner}\ \emph {et~al.}(2014)\citenamefont {Wehner},
  \citenamefont {Reed}, \citenamefont {Li}, \citenamefont {Prabhat},
  \citenamefont {Bacmeister}, \citenamefont {Chen}, \citenamefont {Paciorek},
  \citenamefont {Gleckler}, \citenamefont {Sperber}, \citenamefont {Collins},
  \citenamefont {Gettelman},\ and\ \citenamefont {Jablonowski}}]{wehn14a}%
  \BibitemOpen
  \bibfield  {author} {\bibinfo {author} {\bibfnamefont {M.~F.}\ \bibnamefont
  {Wehner}}, \bibinfo {author} {\bibfnamefont {K.}~\bibnamefont {Reed}},
  \bibinfo {author} {\bibfnamefont {F.}~\bibnamefont {Li}}, \bibinfo {author}
  {\bibnamefont {Prabhat}}, \bibinfo {author} {\bibfnamefont {J.}~\bibnamefont
  {Bacmeister}}, \bibinfo {author} {\bibfnamefont {C.-T.}\ \bibnamefont
  {Chen}}, \bibinfo {author} {\bibfnamefont {C.}~\bibnamefont {Paciorek}},
  \bibinfo {author} {\bibfnamefont {P.}~\bibnamefont {Gleckler}}, \bibinfo
  {author} {\bibfnamefont {K.}~\bibnamefont {Sperber}}, \bibinfo {author}
  {\bibfnamefont {W.~D.}\ \bibnamefont {Collins}}, \bibinfo {author}
  {\bibfnamefont {A.}~\bibnamefont {Gettelman}},\ and\ \bibinfo {author}
  {\bibfnamefont {C.}~\bibnamefont {Jablonowski}},\ }\bibfield  {title}
  {\bibinfo {title} {The effect of horizontal resolution on simulation quality
  in the community atmospheric model, {CAM5.1.}},\ }\href
  {https://doi.org/10.1002/2013MS000276} {\bibfield  {journal} {\bibinfo
  {journal} {J. of Modeling the Earth System}\ }\textbf {\bibinfo {volume}
  {06}},\ \bibinfo {pages} {980} (\bibinfo {year} {2014})}\BibitemShut
  {NoStop}%
\bibitem [{\citenamefont {Sousa}\ \emph {et~al.}(2020)\citenamefont {Sousa},
  \citenamefont {Ramos}, \citenamefont {Raible}, \citenamefont {Messmer},
  \citenamefont {Tom{\'e}}, \citenamefont {Pinto},\ and\ \citenamefont
  {Trigo}}]{sous20a}%
  \BibitemOpen
  \bibfield  {author} {\bibinfo {author} {\bibfnamefont {P.~M.}\ \bibnamefont
  {Sousa}}, \bibinfo {author} {\bibfnamefont {A.~M.}\ \bibnamefont {Ramos}},
  \bibinfo {author} {\bibfnamefont {C.~C.}\ \bibnamefont {Raible}}, \bibinfo
  {author} {\bibfnamefont {M.}~\bibnamefont {Messmer}}, \bibinfo {author}
  {\bibfnamefont {R.}~\bibnamefont {Tom{\'e}}}, \bibinfo {author}
  {\bibfnamefont {J.~G.}\ \bibnamefont {Pinto}},\ and\ \bibinfo {author}
  {\bibfnamefont {R.~M.}\ \bibnamefont {Trigo}},\ }\bibfield  {title} {\bibinfo
  {title} {North {A}tlantic integrated water vapor transport—from 850 to 2100
  {CE}: Impacts on western {E}uropean rainfall},\ }\href@noop {} {\bibfield
  {journal} {\bibinfo  {journal} {J. of Climate}\ }\textbf {\bibinfo {volume}
  {33}},\ \bibinfo {pages} {263} (\bibinfo {year} {2020})}\BibitemShut
  {NoStop}%
\bibitem [{\citenamefont {Rupe}(2022{\natexlab{b}})}]{hurtrac}%
  \BibitemOpen
  \bibfield  {author} {\bibinfo {author} {\bibfnamefont {A.}~\bibnamefont
  {Rupe}},\ }\href@noop {} {\bibinfo {title} {Hurricane tracker video}},\
  \bibinfo {howpublished}
  {\url{https://drive.google.com/file/d/1lWxW0hNVL3eT1VHgOTcCBMdJQrbzrPLK/view?usp=share_link}}
  (\bibinfo {year} {2022}{\natexlab{b}})\BibitemShut {NoStop}%
\bibitem [{\citenamefont {Rupe}(2022{\natexlab{c}})}]{ewetrac}%
  \BibitemOpen
  \bibfield  {author} {\bibinfo {author} {\bibfnamefont {A.}~\bibnamefont
  {Rupe}},\ }\href@noop {} {\bibinfo {title} {General {EWE} tracker video}},\
  \bibinfo {howpublished}
  {\url{https://drive.google.com/file/d/1mFnmHHLxK34IEVvMJA5axPNitVohMy9_/view?usp=share_link}}
  (\bibinfo {year} {2022}{\natexlab{c}})\BibitemShut {NoStop}%
\bibitem [{\citenamefont {O'Brien}\ \emph {et~al.}(2020)\citenamefont
  {O'Brien}, \citenamefont {Risser}, \citenamefont {Loring}, \citenamefont
  {Elbashandy}, \citenamefont {Krishnan}, \citenamefont {Johnson},
  \citenamefont {Patricola}, \citenamefont {O'Brien}, \citenamefont {Mahesh},
  \citenamefont {Prabhat}, \citenamefont {Arriaga~Ramirez}, \citenamefont
  {Rhades}, \citenamefont {Charn}, \citenamefont {Diaz},\ and\ \citenamefont
  {Collins}}]{Obrien20a}%
  \BibitemOpen
  \bibfield  {author} {\bibinfo {author} {\bibfnamefont {T.~A.}\ \bibnamefont
  {O'Brien}}, \bibinfo {author} {\bibfnamefont {M.~D.}\ \bibnamefont {Risser}},
  \bibinfo {author} {\bibfnamefont {B.}~\bibnamefont {Loring}}, \bibinfo
  {author} {\bibfnamefont {A.~A.}\ \bibnamefont {Elbashandy}}, \bibinfo
  {author} {\bibfnamefont {H.}~\bibnamefont {Krishnan}}, \bibinfo {author}
  {\bibfnamefont {J.}~\bibnamefont {Johnson}}, \bibinfo {author} {\bibfnamefont
  {C.~M.}\ \bibnamefont {Patricola}}, \bibinfo {author} {\bibfnamefont {J.~P.}\
  \bibnamefont {O'Brien}}, \bibinfo {author} {\bibfnamefont {A.}~\bibnamefont
  {Mahesh}}, \bibinfo {author} {\bibnamefont {Prabhat}}, \bibinfo {author}
  {\bibfnamefont {S.}~\bibnamefont {Arriaga~Ramirez}}, \bibinfo {author}
  {\bibfnamefont {A.~M.}\ \bibnamefont {Rhades}}, \bibinfo {author}
  {\bibfnamefont {A.}~\bibnamefont {Charn}}, \bibinfo {author} {\bibfnamefont
  {H.~I.}\ \bibnamefont {Diaz}},\ and\ \bibinfo {author} {\bibfnamefont
  {W.~D.}\ \bibnamefont {Collins}},\ }\bibfield  {title} {\bibinfo {title}
  {Detection of atmospheric rivers with inline uncertainty quantification:
  {TECA-BARD} v1.0.1},\ }\href@noop {} {\bibfield  {journal} {\bibinfo
  {journal} {Geoscientific Model Development}\ }\textbf {\bibinfo {volume}
  {13}},\ \bibinfo {pages} {6131} (\bibinfo {year} {2020})}\BibitemShut
  {NoStop}%
\bibitem [{\citenamefont {Ullrich}\ \emph {et~al.}(2021)\citenamefont
  {Ullrich}, \citenamefont {Zarzycki}, \citenamefont {McClenny}, \citenamefont
  {Pinheiro}, \citenamefont {Stansfield},\ and\ \citenamefont
  {Reed}}]{Ullri21a}%
  \BibitemOpen
  \bibfield  {author} {\bibinfo {author} {\bibfnamefont {P.~A.}\ \bibnamefont
  {Ullrich}}, \bibinfo {author} {\bibfnamefont {C.~M.}\ \bibnamefont
  {Zarzycki}}, \bibinfo {author} {\bibfnamefont {E.~E.}\ \bibnamefont
  {McClenny}}, \bibinfo {author} {\bibfnamefont {M.~C.}\ \bibnamefont
  {Pinheiro}}, \bibinfo {author} {\bibfnamefont {A.~M.}\ \bibnamefont
  {Stansfield}},\ and\ \bibinfo {author} {\bibfnamefont {K.~A.}\ \bibnamefont
  {Reed}},\ }\bibfield  {title} {\bibinfo {title} {{TempestExtremes} v2. 1: A
  community framework for feature detection, tracking, and analysis in large
  datasets},\ }\href@noop {} {\bibfield  {journal} {\bibinfo  {journal}
  {Geoscientific Model Development}\ }\textbf {\bibinfo {volume} {14}},\
  \bibinfo {pages} {5023} (\bibinfo {year} {2021})}\BibitemShut {NoStop}%
\bibitem [{\citenamefont {Catto}\ and\ \citenamefont {Pfahl}(2013)}]{Catto13a}%
  \BibitemOpen
  \bibfield  {author} {\bibinfo {author} {\bibfnamefont {J.~L.}\ \bibnamefont
  {Catto}}\ and\ \bibinfo {author} {\bibfnamefont {S.}~\bibnamefont {Pfahl}},\
  }\bibfield  {title} {\bibinfo {title} {The importance of fronts for extreme
  precipitation},\ }\href {https://doi.org/https://doi.org/10.1002/jgrd.50852}
  {\bibfield  {journal} {\bibinfo  {journal} {J. Geophys. Res.: Atmospheres}\
  }\textbf {\bibinfo {volume} {118}},\ \bibinfo {pages} {10,791} (\bibinfo
  {year} {2013})}\BibitemShut {NoStop}%
\bibitem [{\citenamefont {Hodges}(1994)}]{Hodges94a}%
  \BibitemOpen
  \bibfield  {author} {\bibinfo {author} {\bibfnamefont {K.~I.}\ \bibnamefont
  {Hodges}},\ }\bibfield  {title} {\bibinfo {title} {A general method for
  tracking analysis and its application to meteorological data},\ }\href@noop
  {} {\bibfield  {journal} {\bibinfo  {journal} {Monthly Weather Review}\
  }\textbf {\bibinfo {volume} {122}},\ \bibinfo {pages} {2573} (\bibinfo {year}
  {1994})}\BibitemShut {NoStop}%
\bibitem [{\citenamefont {Li}\ \emph {et~al.}(2013)\citenamefont {Li},
  \citenamefont {Collins}, \citenamefont {Wehner},\ and\ \citenamefont
  {Leung}}]{LiF13a}%
  \BibitemOpen
  \bibfield  {author} {\bibinfo {author} {\bibfnamefont {F.}~\bibnamefont
  {Li}}, \bibinfo {author} {\bibfnamefont {W.~D.}\ \bibnamefont {Collins}},
  \bibinfo {author} {\bibfnamefont {M.~F.}\ \bibnamefont {Wehner}},\ and\
  \bibinfo {author} {\bibfnamefont {L.~R.}\ \bibnamefont {Leung}},\ }\bibfield
  {title} {\bibinfo {title} {Hurricanes in an aquaplanet world: Implications of
  the impacts of external forcing and model horizontal resolution},\ }\href
  {https://doi.org/https://doi.org/10.1002/jame.20020} {\bibfield  {journal}
  {\bibinfo  {journal} {Journal of Advances in Modeling Earth Systems}\
  }\textbf {\bibinfo {volume} {5}},\ \bibinfo {pages} {134} (\bibinfo {year}
  {2013})}\BibitemShut {NoStop}%
\bibitem [{\citenamefont {Berry}\ \emph {et~al.}(2020)\citenamefont {Berry},
  \citenamefont {Giannakis},\ and\ \citenamefont {Harlim}}]{berr20a}%
  \BibitemOpen
  \bibfield  {author} {\bibinfo {author} {\bibfnamefont {T.}~\bibnamefont
  {Berry}}, \bibinfo {author} {\bibfnamefont {D.}~\bibnamefont {Giannakis}},\
  and\ \bibinfo {author} {\bibfnamefont {J.}~\bibnamefont {Harlim}},\
  }\bibfield  {title} {\bibinfo {title} {Bridging data science and dynamical
  systems theory},\ }\href@noop {} {\bibfield  {journal} {\bibinfo  {journal}
  {Notices of the American Mathematical Society}\ }\textbf {\bibinfo {volume}
  {67}},\ \bibinfo {pages} {1336} (\bibinfo {year} {2020})}\BibitemShut
  {NoStop}%
\bibitem [{\citenamefont {Rupe}\ \emph {et~al.}(2022)\citenamefont {Rupe},
  \citenamefont {Vesselinov},\ and\ \citenamefont {Crutchfield}}]{rupe22b}%
  \BibitemOpen
  \bibfield  {author} {\bibinfo {author} {\bibfnamefont {A.}~\bibnamefont
  {Rupe}}, \bibinfo {author} {\bibfnamefont {V.~V.}\ \bibnamefont
  {Vesselinov}},\ and\ \bibinfo {author} {\bibfnamefont {J.~P.}\ \bibnamefont
  {Crutchfield}},\ }\bibfield  {title} {\bibinfo {title} {Nonequilibrium
  statistical mechanics and optimal prediction of partially-observed complex
  systems},\ }\href {https://doi.org/10.1088/1367-2630/ac95b7} {\bibfield
  {journal} {\bibinfo  {journal} {New Journal of Phys.}\ }\textbf {\bibinfo
  {volume} {24}},\ \bibinfo {pages} {103033} (\bibinfo {year}
  {2022})}\BibitemShut {NoStop}%
\bibitem [{\citenamefont {Runge}\ \emph {et~al.}(2015)\citenamefont {Runge},
  \citenamefont {Petoukhov}, \citenamefont {Donges}, \citenamefont {Hlinka},
  \citenamefont {Jajcay}, \citenamefont {Vejmelka}, \citenamefont {Hartman},
  \citenamefont {Marwan}, \citenamefont {Palu{\v{s}}},\ and\ \citenamefont
  {Kurths}}]{rung15a}%
  \BibitemOpen
  \bibfield  {author} {\bibinfo {author} {\bibfnamefont {J.}~\bibnamefont
  {Runge}}, \bibinfo {author} {\bibfnamefont {V.}~\bibnamefont {Petoukhov}},
  \bibinfo {author} {\bibfnamefont {J.~F.}\ \bibnamefont {Donges}}, \bibinfo
  {author} {\bibfnamefont {J.}~\bibnamefont {Hlinka}}, \bibinfo {author}
  {\bibfnamefont {N.}~\bibnamefont {Jajcay}}, \bibinfo {author} {\bibfnamefont
  {M.}~\bibnamefont {Vejmelka}}, \bibinfo {author} {\bibfnamefont
  {D.}~\bibnamefont {Hartman}}, \bibinfo {author} {\bibfnamefont
  {N.}~\bibnamefont {Marwan}}, \bibinfo {author} {\bibfnamefont
  {M.}~\bibnamefont {Palu{\v{s}}}},\ and\ \bibinfo {author} {\bibfnamefont
  {J.}~\bibnamefont {Kurths}},\ }\bibfield  {title} {\bibinfo {title}
  {Identifying causal gateways and mediators in complex spatio-temporal
  systems},\ }\href@noop {} {\bibfield  {journal} {\bibinfo  {journal} {Nature
  Comms.}\ }\textbf {\bibinfo {volume} {6}},\ \bibinfo {pages} {1} (\bibinfo
  {year} {2015})}\BibitemShut {NoStop}%
\bibitem [{\citenamefont {Klus}\ \emph {et~al.}(2019)\citenamefont {Klus},
  \citenamefont {Husic}, \citenamefont {Mollenhauer},\ and\ \citenamefont
  {No\'{e}}}]{Klus19a}%
  \BibitemOpen
  \bibfield  {author} {\bibinfo {author} {\bibfnamefont {S.}~\bibnamefont
  {Klus}}, \bibinfo {author} {\bibfnamefont {B.~E.}\ \bibnamefont {Husic}},
  \bibinfo {author} {\bibfnamefont {M.}~\bibnamefont {Mollenhauer}},\ and\
  \bibinfo {author} {\bibfnamefont {F.}~\bibnamefont {No\'{e}}},\ }\bibfield
  {title} {\bibinfo {title} {Kernel methods for detecting coherent structures
  in dynamical data},\ }\href {https://doi.org/10.1063/1.5100267} {\bibfield
  {journal} {\bibinfo  {journal} {Chaos}\ }\textbf {\bibinfo {volume} {29}},\
  \bibinfo {pages} {123112} (\bibinfo {year} {2019})}\BibitemShut {NoStop}%
\bibitem [{\citenamefont {Ralph}\ \emph {et~al.}(2019)\citenamefont {Ralph},
  \citenamefont {Rutz}, \citenamefont {Cordeira}, \citenamefont {Dettinger},
  \citenamefont {Anderson}, \citenamefont {Reynolds}, \citenamefont {Schick},\
  and\ \citenamefont {Smallcomb}}]{ralph19a}%
  \BibitemOpen
  \bibfield  {author} {\bibinfo {author} {\bibfnamefont {F.~M.}\ \bibnamefont
  {Ralph}}, \bibinfo {author} {\bibfnamefont {J.~J.}\ \bibnamefont {Rutz}},
  \bibinfo {author} {\bibfnamefont {J.~M.}\ \bibnamefont {Cordeira}}, \bibinfo
  {author} {\bibfnamefont {M.}~\bibnamefont {Dettinger}}, \bibinfo {author}
  {\bibfnamefont {M.}~\bibnamefont {Anderson}}, \bibinfo {author}
  {\bibfnamefont {D.}~\bibnamefont {Reynolds}}, \bibinfo {author}
  {\bibfnamefont {L.~J.}\ \bibnamefont {Schick}},\ and\ \bibinfo {author}
  {\bibfnamefont {C.}~\bibnamefont {Smallcomb}},\ }\bibfield  {title} {\bibinfo
  {title} {A scale to characterize the strength and impacts of atmospheric
  rivers},\ }\href@noop {} {\bibfield  {journal} {\bibinfo  {journal} {Bulletin
  of the American Meteorological Society}\ }\textbf {\bibinfo {volume} {100}},\
  \bibinfo {pages} {269} (\bibinfo {year} {2019})}\BibitemShut {NoStop}%
\bibitem [{\citenamefont {Wetzel}\ \emph {et~al.}(2017)\citenamefont {Wetzel},
  \citenamefont {Smith},\ and\ \citenamefont {Stechmann}}]{wetz17a}%
  \BibitemOpen
  \bibfield  {author} {\bibinfo {author} {\bibfnamefont {A.~N.}\ \bibnamefont
  {Wetzel}}, \bibinfo {author} {\bibfnamefont {L.~M.}\ \bibnamefont {Smith}},\
  and\ \bibinfo {author} {\bibfnamefont {S.~N.}\ \bibnamefont {Stechmann}},\
  }\bibfield  {title} {\bibinfo {title} {Moisture transport due to baroclinic
  waves: Linear analysis of precipitating quasi-geostrophic dynamics},\
  }\href@noop {} {\bibfield  {journal} {\bibinfo  {journal} {Mathematics of
  Climate and Weather Forecasting}\ }\textbf {\bibinfo {volume} {3}},\ \bibinfo
  {pages} {28} (\bibinfo {year} {2017})}\BibitemShut {NoStop}%
\bibitem [{\citenamefont {Weeks}\ \emph {et~al.}(1996)\citenamefont {Weeks},
  \citenamefont {Urbach},\ and\ \citenamefont {Swinney}}]{Week96a}%
  \BibitemOpen
  \bibfield  {author} {\bibinfo {author} {\bibfnamefont {E.~R.}\ \bibnamefont
  {Weeks}}, \bibinfo {author} {\bibfnamefont {J.~S.}\ \bibnamefont {Urbach}},\
  and\ \bibinfo {author} {\bibfnamefont {H.~L.}\ \bibnamefont {Swinney}},\
  }\bibfield  {title} {\bibinfo {title} {Anomalous diffusion in asymmetric
  random walks with a quasi-geostrophic flow example},\ }\href@noop {}
  {\bibfield  {journal} {\bibinfo  {journal} {Physica D}\ }\textbf {\bibinfo
  {volume} {97}},\ \bibinfo {pages} {291} (\bibinfo {year} {1996})}\BibitemShut
  {NoStop}%
\bibitem [{\citenamefont {Taira}\ \emph {et~al.}(2017)\citenamefont {Taira},
  \citenamefont {Brunton}, \citenamefont {Dawson}, \citenamefont {Rowley},
  \citenamefont {Colonius}, \citenamefont {McKeon}, \citenamefont {Schmidt},
  \citenamefont {Gordeyev}, \citenamefont {Theofilis},\ and\ \citenamefont
  {Ukeiley}}]{tair17a}%
  \BibitemOpen
  \bibfield  {author} {\bibinfo {author} {\bibfnamefont {K.}~\bibnamefont
  {Taira}}, \bibinfo {author} {\bibfnamefont {S.~L.}\ \bibnamefont {Brunton}},
  \bibinfo {author} {\bibfnamefont {S.~T.}\ \bibnamefont {Dawson}}, \bibinfo
  {author} {\bibfnamefont {C.~W.}\ \bibnamefont {Rowley}}, \bibinfo {author}
  {\bibfnamefont {T.}~\bibnamefont {Colonius}}, \bibinfo {author}
  {\bibfnamefont {B.~J.}\ \bibnamefont {McKeon}}, \bibinfo {author}
  {\bibfnamefont {O.~T.}\ \bibnamefont {Schmidt}}, \bibinfo {author}
  {\bibfnamefont {S.}~\bibnamefont {Gordeyev}}, \bibinfo {author}
  {\bibfnamefont {V.}~\bibnamefont {Theofilis}},\ and\ \bibinfo {author}
  {\bibfnamefont {L.~S.}\ \bibnamefont {Ukeiley}},\ }\bibfield  {title}
  {\bibinfo {title} {Modal analysis of fluid flows: An overview},\ }\href@noop
  {} {\bibfield  {journal} {\bibinfo  {journal} {AIAA Journal}\ }\textbf
  {\bibinfo {volume} {55}},\ \bibinfo {pages} {4013} (\bibinfo {year}
  {2017})}\BibitemShut {NoStop}%
\bibitem [{\citenamefont {Tu}\ \emph {et~al.}(2014)\citenamefont {Tu},
  \citenamefont {Rowley}, \citenamefont {Luchtenburg}, \citenamefont
  {Brunton},\ and\ \citenamefont {Kutz}}]{Tu14a}%
  \BibitemOpen
  \bibfield  {author} {\bibinfo {author} {\bibfnamefont {J.~H.}\ \bibnamefont
  {Tu}}, \bibinfo {author} {\bibfnamefont {C.~W.}\ \bibnamefont {Rowley}},
  \bibinfo {author} {\bibfnamefont {D.~M.}\ \bibnamefont {Luchtenburg}},
  \bibinfo {author} {\bibfnamefont {S.~L.}\ \bibnamefont {Brunton}},\ and\
  \bibinfo {author} {\bibfnamefont {J.~N.}\ \bibnamefont {Kutz}},\ }\bibfield
  {title} {\bibinfo {title} {On dynamic mode decomposition: Theory and
  applications},\ }\href@noop {} {\bibfield  {journal} {\bibinfo  {journal} {J.
  Comp. Dyn.}\ }\textbf {\bibinfo {volume} {1}},\ \bibinfo {pages} {391}
  (\bibinfo {year} {2014})}\BibitemShut {NoStop}%
\bibitem [{\citenamefont {Shalizi}(2003)}]{Shal03a}%
  \BibitemOpen
  \bibfield  {author} {\bibinfo {author} {\bibfnamefont {C.}~\bibnamefont
  {Shalizi}},\ }\bibfield  {title} {\bibinfo {title} {Optimal nonlinear
  prediction of random fields on networks},\ }\href@noop {} {\bibfield
  {journal} {\bibinfo  {journal} {DMTCS Proceedings}\ }\textbf {\bibinfo
  {volume} {vol. AB}},\ \bibinfo {pages} {11} (\bibinfo {year}
  {2003})}\BibitemShut {NoStop}%
\bibitem [{\citenamefont {Caires}\ and\ \citenamefont
  {Ferreira}(2005)}]{cair05a}%
  \BibitemOpen
  \bibfield  {author} {\bibinfo {author} {\bibfnamefont {S.}~\bibnamefont
  {Caires}}\ and\ \bibinfo {author} {\bibfnamefont {J.~A.}\ \bibnamefont
  {Ferreira}},\ }\bibfield  {title} {\bibinfo {title} {On the non-parametric
  prediction of conditionally stationary sequences},\ }\href@noop {} {\bibfield
   {journal} {\bibinfo  {journal} {Statistical Inference for Stochastic
  Processes}\ }\textbf {\bibinfo {volume} {8}},\ \bibinfo {pages} {151}
  (\bibinfo {year} {2005})}\BibitemShut {NoStop}%
\bibitem [{\citenamefont {Bronstein}\ \emph {et~al.}(2017)\citenamefont
  {Bronstein}, \citenamefont {Bruna}, \citenamefont {LeCun}, \citenamefont
  {Szlam},\ and\ \citenamefont {Vandergheynst}}]{bron17a}%
  \BibitemOpen
  \bibfield  {author} {\bibinfo {author} {\bibfnamefont {M.~M.}\ \bibnamefont
  {Bronstein}}, \bibinfo {author} {\bibfnamefont {J.}~\bibnamefont {Bruna}},
  \bibinfo {author} {\bibfnamefont {Y.}~\bibnamefont {LeCun}}, \bibinfo
  {author} {\bibfnamefont {A.}~\bibnamefont {Szlam}},\ and\ \bibinfo {author}
  {\bibfnamefont {P.}~\bibnamefont {Vandergheynst}},\ }\bibfield  {title}
  {\bibinfo {title} {Geometric deep learning: going beyond {E}uclidean data},\
  }\href@noop {} {\bibfield  {journal} {\bibinfo  {journal} {IEEE Sig. Proc.
  Mag.}\ }\textbf {\bibinfo {volume} {34}},\ \bibinfo {pages} {18} (\bibinfo
  {year} {2017})}\BibitemShut {NoStop}%
\bibitem [{\citenamefont {Bronstein}\ \emph {et~al.}(2021)\citenamefont
  {Bronstein}, \citenamefont {Bruna}, \citenamefont {Cohen},\ and\
  \citenamefont {Veli{\v{c}}kovi{\'c}}}]{bron21a}%
  \BibitemOpen
  \bibfield  {author} {\bibinfo {author} {\bibfnamefont {M.~M.}\ \bibnamefont
  {Bronstein}}, \bibinfo {author} {\bibfnamefont {J.}~\bibnamefont {Bruna}},
  \bibinfo {author} {\bibfnamefont {T.}~\bibnamefont {Cohen}},\ and\ \bibinfo
  {author} {\bibfnamefont {P.}~\bibnamefont {Veli{\v{c}}kovi{\'c}}},\
  }\bibfield  {title} {\bibinfo {title} {Geometric deep learning: Grids,
  groups, graphs, geodesics, and gauges},\ }\href@noop {} {\bibfield  {journal}
  {\bibinfo  {journal} {arXiv preprint arXiv:2104.13478}\ } (\bibinfo {year}
  {2021})}\BibitemShut {NoStop}%
\bibitem [{\citenamefont {Goerg}\ and\ \citenamefont {Shalizi}()}]{Goer12a}%
  \BibitemOpen
  \bibfield  {author} {\bibinfo {author} {\bibfnamefont {G.}~\bibnamefont
  {Goerg}}\ and\ \bibinfo {author} {\bibfnamefont {C.}~\bibnamefont
  {Shalizi}},\ }\bibfield  {title} {\bibinfo {title} {{LICORS}: Light cone
  reconstruction of states for non-parametric forecasting of spatio-temporal
  systems},\ }\href@noop {} {\bibinfo  {journal} {arXiv:1206.2398}\
  }\BibitemShut {NoStop}%
\bibitem [{\citenamefont {J{\"a}nicke}\ \emph {et~al.}(2007)\citenamefont
  {J{\"a}nicke}, \citenamefont {Wiebel}, \citenamefont {Scheuermann},\ and\
  \citenamefont {Kollmann}}]{Jani07a}%
  \BibitemOpen
\bibfield  {journal} {  }\bibfield  {author} {\bibinfo {author} {\bibfnamefont
  {H.}~\bibnamefont {J{\"a}nicke}}, \bibinfo {author} {\bibfnamefont
  {A.}~\bibnamefont {Wiebel}}, \bibinfo {author} {\bibfnamefont
  {G.}~\bibnamefont {Scheuermann}},\ and\ \bibinfo {author} {\bibfnamefont
  {W.}~\bibnamefont {Kollmann}},\ }\bibfield  {title} {\bibinfo {title}
  {Multifield visualization using local statistical complexity},\ }\href@noop
  {} {\bibfield  {journal} {\bibinfo  {journal} {IEEE Trans. Vis. Comp.
  Graphics}\ }\textbf {\bibinfo {volume} {13}},\ \bibinfo {pages} {1384}
  (\bibinfo {year} {2007})}\BibitemShut {NoStop}%
\bibitem [{\citenamefont {Ester}\ \emph {et~al.}(1996)\citenamefont {Ester},
  \citenamefont {Kriegel}, \citenamefont {Sander}, \citenamefont {Xu} \emph
  {et~al.}}]{dbscan}%
  \BibitemOpen
  \bibfield  {author} {\bibinfo {author} {\bibfnamefont {M.}~\bibnamefont
  {Ester}}, \bibinfo {author} {\bibfnamefont {H.-P.}\ \bibnamefont {Kriegel}},
  \bibinfo {author} {\bibfnamefont {J.}~\bibnamefont {Sander}}, \bibinfo
  {author} {\bibfnamefont {X.}~\bibnamefont {Xu}}, \emph {et~al.},\ }\bibfield
  {title} {\bibinfo {title} {A density-based algorithm for discovering clusters
  in large spatial databases with noise.},\ }in\ \href@noop {} {\emph {\bibinfo
  {booktitle} {KDD}}},\ Vol.~\bibinfo {volume} {96}\ (\bibinfo {year} {1996})\
  pp.\ \bibinfo {pages} {226--231}\BibitemShut {NoStop}%
\bibitem [{\citenamefont {Balestriero}\ and\ \citenamefont
  {Baraniuk}(2021)}]{bale18a}%
  \BibitemOpen
  \bibfield  {author} {\bibinfo {author} {\bibfnamefont {R.}~\bibnamefont
  {Balestriero}}\ and\ \bibinfo {author} {\bibfnamefont {R.~G.}\ \bibnamefont
  {Baraniuk}},\ }\bibfield  {title} {\bibinfo {title} {Mad {M}ax: Affine spline
  insights into deep learning},\ }\href
  {https://doi.org/10.1109/JPROC.2020.3042100} {\bibfield  {journal} {\bibinfo
  {journal} {Proceedings of the IEEE}\ }\textbf {\bibinfo {volume} {109}},\
  \bibinfo {pages} {704} (\bibinfo {year} {2021})}\BibitemShut {NoStop}%
\bibitem [{\citenamefont {Packard}\ \emph {et~al.}(1980)\citenamefont
  {Packard}, \citenamefont {Crutchfield}, \citenamefont {Farmer},\ and\
  \citenamefont {Shaw}}]{Pack80}%
  \BibitemOpen
  \bibfield  {author} {\bibinfo {author} {\bibfnamefont {N.~H.}\ \bibnamefont
  {Packard}}, \bibinfo {author} {\bibfnamefont {J.~P.}\ \bibnamefont
  {Crutchfield}}, \bibinfo {author} {\bibfnamefont {J.~D.}\ \bibnamefont
  {Farmer}},\ and\ \bibinfo {author} {\bibfnamefont {R.~S.}\ \bibnamefont
  {Shaw}},\ }\bibfield  {title} {\bibinfo {title} {Geometry from a time
  series},\ }\href@noop {} {\bibfield  {journal} {\bibinfo  {journal} {Phys.
  Rev. Let.}\ }\textbf {\bibinfo {volume} {45}},\ \bibinfo {pages} {712}
  (\bibinfo {year} {1980})}\BibitemShut {NoStop}%
\bibitem [{\citenamefont {B{\'e}nard}(1901)}]{Bena01a}%
  \BibitemOpen
  \bibfield  {author} {\bibinfo {author} {\bibfnamefont {H.}~\bibnamefont
  {B{\'e}nard}},\ }\href@noop {} {\emph {\bibinfo {title} {Les Tourbillons
  Cellulaires dans une nappe Liquide Propageant de la Chaleur par Convection:
  en R{\'e}gime Permanent}}}\ (\bibinfo  {publisher} {Gauthier-Villars},\
  \bibinfo {year} {1901})\BibitemShut {NoStop}%
\bibitem [{\citenamefont {Rayleigh}(1916)}]{Rayl16a}%
  \BibitemOpen
  \bibfield  {author} {\bibinfo {author} {\bibfnamefont {L.}~\bibnamefont
  {Rayleigh}},\ }\bibfield  {title} {\bibinfo {title} {On convection currents
  in a horizontal layer of fluid, when the higher temperature is on the under
  side},\ }\href@noop {} {\bibfield  {journal} {\bibinfo  {journal} {Phil. Mag.
  (Series 6)}\ }\textbf {\bibinfo {volume} {32}},\ \bibinfo {pages} {529}
  (\bibinfo {year} {1916})}\BibitemShut {NoStop}%
\bibitem [{\citenamefont {Chandrasekhar}(1968)}]{Chan68a}%
  \BibitemOpen
  \bibfield  {author} {\bibinfo {author} {\bibfnamefont {S.}~\bibnamefont
  {Chandrasekhar}},\ }\href@noop {} {\emph {\bibinfo {title} {Hydrodynamic and
  Hydromagnetic Stability}}}\ (\bibinfo  {publisher} {Oxford, Clarendon
  Press},\ \bibinfo {year} {1968})\BibitemShut {NoStop}%
\bibitem [{\citenamefont {Busse}(1978)}]{Buss78a}%
  \BibitemOpen
  \bibfield  {author} {\bibinfo {author} {\bibfnamefont {F.~H.}\ \bibnamefont
  {Busse}},\ }\bibfield  {title} {\bibinfo {title} {Non-linear properties of
  thermal convection},\ }\href@noop {} {\bibfield  {journal} {\bibinfo
  {journal} {Reports on Progress in Physics}\ }\textbf {\bibinfo {volume}
  {41}},\ \bibinfo {pages} {1929} (\bibinfo {year} {1978})}\BibitemShut
  {NoStop}%
\bibitem [{\citenamefont {Fenstermacher}\ \emph {et~al.}(1979)\citenamefont
  {Fenstermacher}, \citenamefont {Swinney},\ and\ \citenamefont
  {Gollub}}]{Fens79a}%
  \BibitemOpen
  \bibfield  {author} {\bibinfo {author} {\bibfnamefont {P.}~\bibnamefont
  {Fenstermacher}}, \bibinfo {author} {\bibfnamefont {H.}~\bibnamefont
  {Swinney}},\ and\ \bibinfo {author} {\bibfnamefont {J.}~\bibnamefont
  {Gollub}},\ }\bibfield  {title} {\bibinfo {title} {Dynamical instabilities
  and the transition to chaotic {T}aylor vortex flow},\ }\href@noop {}
  {\bibfield  {journal} {\bibinfo  {journal} {J. Fluid Mech.}\ }\textbf
  {\bibinfo {volume} {94}},\ \bibinfo {pages} {103} (\bibinfo {year}
  {1979})}\BibitemShut {NoStop}%
\bibitem [{\citenamefont {Steinberg}\ \emph {et~al.}(1985)\citenamefont
  {Steinberg}, \citenamefont {Ahlers},\ and\ \citenamefont {Cannell}}]{Stei85}%
  \BibitemOpen
  \bibfield  {author} {\bibinfo {author} {\bibfnamefont {V.}~\bibnamefont
  {Steinberg}}, \bibinfo {author} {\bibfnamefont {G.}~\bibnamefont {Ahlers}},\
  and\ \bibinfo {author} {\bibfnamefont {D.~S.}\ \bibnamefont {Cannell}},\
  }\bibfield  {title} {\bibinfo {title} {Pattern formation and wave-number
  selection by {R}ayleigh-{B}{\'e}nard convection in a cylindrical container},\
  }\href@noop {} {\bibfield  {journal} {\bibinfo  {journal} {Physica Scripta}\
  }\textbf {\bibinfo {volume} {T9}},\ \bibinfo {pages} {97} (\bibinfo {year}
  {1985})}\BibitemShut {NoStop}%
\bibitem [{\citenamefont {Cross}\ and\ \citenamefont
  {Hohenberg}(1993)}]{Cros93a}%
  \BibitemOpen
  \bibfield  {author} {\bibinfo {author} {\bibfnamefont {M.~C.}\ \bibnamefont
  {Cross}}\ and\ \bibinfo {author} {\bibfnamefont {P.~C.}\ \bibnamefont
  {Hohenberg}},\ }\bibfield  {title} {\bibinfo {title} {Pattern formation
  outside of equilibrium},\ }\href@noop {} {\bibfield  {journal} {\bibinfo
  {journal} {Rev. Mod. Phys.}\ }\textbf {\bibinfo {volume} {65}},\ \bibinfo
  {pages} {851} (\bibinfo {year} {1993})}\BibitemShut {NoStop}%
\bibitem [{\citenamefont {Heisenberg}(1967)}]{Heis67a}%
  \BibitemOpen
  \bibfield  {author} {\bibinfo {author} {\bibfnamefont {W.}~\bibnamefont
  {Heisenberg}},\ }\bibfield  {title} {\bibinfo {title} {Nonlinear problems in
  physics},\ }\href@noop {} {\bibfield  {journal} {\bibinfo  {journal} {Physics
  Today}\ }\textbf {\bibinfo {volume} {20}},\ \bibinfo {pages} {23} (\bibinfo
  {year} {1967})}\BibitemShut {NoStop}%
\bibitem [{\citenamefont {Liu}(1988)}]{liu88a}%
  \BibitemOpen
  \bibfield  {author} {\bibinfo {author} {\bibfnamefont {J.~T.~C.}\
  \bibnamefont {Liu}},\ }\bibfield  {title} {\bibinfo {title} {Contributions to
  the understanding of large-scale coherent structures in developing free
  turbulent shear flows},\ }\href@noop {} {\bibfield  {journal} {\bibinfo
  {journal} {Advances in applied mechanics}\ }\textbf {\bibinfo {volume}
  {26}},\ \bibinfo {pages} {183} (\bibinfo {year} {1988})}\BibitemShut
  {NoStop}%
\bibitem [{\citenamefont {Liepmann}(1952)}]{liep52a}%
  \BibitemOpen
  \bibfield  {author} {\bibinfo {author} {\bibfnamefont {H.~W.}\ \bibnamefont
  {Liepmann}},\ }\bibfield  {title} {\bibinfo {title} {Aspects of the
  turbulence problem},\ }\href@noop {} {\bibfield  {journal} {\bibinfo
  {journal} {Zeitschrift f{\"u}r angewandte Mathematik und Physik ZAMP}\
  }\textbf {\bibinfo {volume} {3}},\ \bibinfo {pages} {407} (\bibinfo {year}
  {1952})}\BibitemShut {NoStop}%
\bibitem [{\citenamefont {Townsend}(1956)}]{town56a}%
  \BibitemOpen
  \bibfield  {author} {\bibinfo {author} {\bibfnamefont {A.~A.}\ \bibnamefont
  {Townsend}},\ }\href@noop {} {\emph {\bibinfo {title} {The structure of
  turbulent shear flow}}}\ (\bibinfo  {publisher} {Cambridge university
  press},\ \bibinfo {year} {1956})\BibitemShut {NoStop}%
\bibitem [{\citenamefont {Lumley}(1967)}]{luml67a}%
  \BibitemOpen
  \bibfield  {author} {\bibinfo {author} {\bibfnamefont {J.~L.}\ \bibnamefont
  {Lumley}},\ }\bibfield  {title} {\bibinfo {title} {The structure of
  inhomogeneous turbulence},\ }\href@noop {} {\bibfield  {journal} {\bibinfo
  {journal} {Atmospheric turbulence and radio wave propagation}\ ,\ \bibinfo
  {pages} {166}} (\bibinfo {year} {1967})}\BibitemShut {NoStop}%
\bibitem [{\citenamefont {McWilliams}(1983)}]{mcwil83a}%
  \BibitemOpen
  \bibfield  {author} {\bibinfo {author} {\bibfnamefont {J.~C.}\ \bibnamefont
  {McWilliams}},\ }\bibfield  {title} {\bibinfo {title} {On the relevance of
  two-dimensional turbulence to geophysical fluid motions},\ }\href@noop {}
  {\bibfield  {journal} {\bibinfo  {journal} {Journal de Mecanique Theorique et
  Appliquee Supplement}\ ,\ \bibinfo {pages} {83}} (\bibinfo {year}
  {1983})}\BibitemShut {NoStop}%
\bibitem [{\citenamefont {Parker}(2016)}]{park16a}%
  \BibitemOpen
  \bibfield  {author} {\bibinfo {author} {\bibfnamefont {W.~S.}\ \bibnamefont
  {Parker}},\ }\bibfield  {title} {\bibinfo {title} {Reanalyses and
  observations: What’s the difference?},\ }\href@noop {} {\bibfield
  {journal} {\bibinfo  {journal} {Bulletin of the American Meteorological
  Society}\ }\textbf {\bibinfo {volume} {97}},\ \bibinfo {pages} {1565}
  (\bibinfo {year} {2016})}\BibitemShut {NoStop}%
\bibitem [{\citenamefont {Dagon}\ \emph {et~al.}(2022)\citenamefont {Dagon},
  \citenamefont {Truesdale}, \citenamefont {Biard}, \citenamefont {Kunkel},
  \citenamefont {Meehl},\ and\ \citenamefont {Molina}}]{dagon22a}%
  \BibitemOpen
  \bibfield  {author} {\bibinfo {author} {\bibfnamefont {K.}~\bibnamefont
  {Dagon}}, \bibinfo {author} {\bibfnamefont {J.}~\bibnamefont {Truesdale}},
  \bibinfo {author} {\bibfnamefont {J.~C.}\ \bibnamefont {Biard}}, \bibinfo
  {author} {\bibfnamefont {K.~E.}\ \bibnamefont {Kunkel}}, \bibinfo {author}
  {\bibfnamefont {G.~A.}\ \bibnamefont {Meehl}},\ and\ \bibinfo {author}
  {\bibfnamefont {M.~J.}\ \bibnamefont {Molina}},\ }\bibfield  {title}
  {\bibinfo {title} {Machine learning-based detection of weather fronts and
  associated extreme precipitation in historical and future climates},\ }\href
  {https://doi.org/https://doi.org/10.1029/2022JD037038} {\bibfield  {journal}
  {\bibinfo  {journal} {Journal of Geophysical Research: Atmospheres}\ }\textbf
  {\bibinfo {volume} {127}},\ \bibinfo {pages} {e2022JD037038} (\bibinfo {year}
  {2022})}\BibitemShut {NoStop}%
\bibitem [{\citenamefont {Knutson}\ \emph {et~al.}(2007)\citenamefont
  {Knutson}, \citenamefont {Sirutis}, \citenamefont {Garner}, \citenamefont
  {Held},\ and\ \citenamefont {Tuleya}}]{Knut07a}%
  \BibitemOpen
  \bibfield  {author} {\bibinfo {author} {\bibfnamefont {T.~R.}\ \bibnamefont
  {Knutson}}, \bibinfo {author} {\bibfnamefont {J.~J.}\ \bibnamefont
  {Sirutis}}, \bibinfo {author} {\bibfnamefont {S.~T.}\ \bibnamefont {Garner}},
  \bibinfo {author} {\bibfnamefont {I.~M.}\ \bibnamefont {Held}},\ and\
  \bibinfo {author} {\bibfnamefont {R.~E.}\ \bibnamefont {Tuleya}},\ }\bibfield
   {title} {\bibinfo {title} {Simulation of the recent multidecadal increase of
  atlantic hurricane activity using an 18-km-grid regional model},\ }\href@noop
  {} {\bibfield  {journal} {\bibinfo  {journal} {Bulletin of the American
  Meteorological Society}\ }\textbf {\bibinfo {volume} {88}},\ \bibinfo {pages}
  {1549} (\bibinfo {year} {2007})}\BibitemShut {NoStop}%
\bibitem [{\citenamefont {Chavas}\ \emph {et~al.}(2015)\citenamefont {Chavas},
  \citenamefont {Lin},\ and\ \citenamefont {Emanuel}}]{Chava15a}%
  \BibitemOpen
  \bibfield  {author} {\bibinfo {author} {\bibfnamefont {D.~R.}\ \bibnamefont
  {Chavas}}, \bibinfo {author} {\bibfnamefont {N.}~\bibnamefont {Lin}},\ and\
  \bibinfo {author} {\bibfnamefont {K.}~\bibnamefont {Emanuel}},\ }\bibfield
  {title} {\bibinfo {title} {A model for the complete radial structure of the
  tropical cyclone wind field. {P}art i: Comparison with observed structure},\
  }\href@noop {} {\bibfield  {journal} {\bibinfo  {journal} {Journal of the
  Atmospheric Sciences}\ }\textbf {\bibinfo {volume} {72}},\ \bibinfo {pages}
  {3647} (\bibinfo {year} {2015})}\BibitemShut {NoStop}%
\bibitem [{\citenamefont {NASA}(3 14)}]{cassini}%
  \BibitemOpen
  \bibfield  {author} {\bibinfo {author} {\bibnamefont {NASA}},\ }\href@noop {}
  {\bibinfo {title} {Jupiter cloud sequence from cassini}},\ \bibinfo
  {howpublished} {\url{https://svs.gsfc.nasa.gov/cgi-bin/details.cgi?aid=3610}}
  (\bibinfo {year} {Accessed: 2019-03-14})\BibitemShut {NoStop}%
\bibitem [{\citenamefont {Hadjighasem}\ and\ \citenamefont
  {Haller}(2016)}]{Hadj16a}%
  \BibitemOpen
  \bibfield  {author} {\bibinfo {author} {\bibfnamefont {A.}~\bibnamefont
  {Hadjighasem}}\ and\ \bibinfo {author} {\bibfnamefont {G.}~\bibnamefont
  {Haller}},\ }\bibfield  {title} {\bibinfo {title} {Geodesic transport
  barriers in {J}upiter's atmosphere: Video-based analysis},\ }\href@noop {}
  {\bibfield  {journal} {\bibinfo  {journal} {SIAM Review}\ }\textbf {\bibinfo
  {volume} {58}},\ \bibinfo {pages} {69} (\bibinfo {year} {2016})}\BibitemShut
  {NoStop}%
\end{thebibliography}%

\pagebreak
\clearpage
\onecolumngrid
\begin{center}
\textbf{\large Supplementary Information}
\end{center}
\setcounter{equation}{0}
\setcounter{figure}{0}
\setcounter{table}{0}
\setcounter{page}{1}
\makeatletter
\renewcommand{\theequation}{S\arabic{equation}}
\renewcommand{\thefigure}{S\arabic{figure}}

\subsection{Methodology}

\subsubsection{Spatiotemporal Systems}

Local causal states are learned representations that extract emergent
organization from spatially-extended dynamical systems. The \emph{behaviors} of
spatiotemporal systems are \emph{spacetime fields}, elements of the extended
phase space that includes time as a dimension. Mathematically, a spacetime
field is a tensor $X(\mathbf{r}, t): \mathbb{S} \rightarrow \mathbb{R}^n$,
where $\mathbb{S}$ is some spacetime coordinate system and $n$ is the number of
physical field quantities---e.g., temperature, pressure, velocity components,
and so on. Each coordinate in spacetime maps to a length-$n$ vector of the
values of the physical quantities at that point. 

The systems presently analyzed live in a simple Euclidean spacetime, so that
$\mathbb{S} = \mathbb{R}^{d+1}$, where $d$ is the number of spatial dimensions.
Note though that the spacetime data from the climate reanalysis and Jupiter's
clouds are Euclidean projections from a spherical spatial coordinate system. 

Numerical data, with which any data-driven method must necessarily work, will
be given on some finite grid $\mathbb{G}$ of dimension $(X, Y, T)$ for two
spatial dimensions. The data used here is on a simple integer Euclidean grid
(as opposed to a complex adaptive mesh, or spherical grid), so that for each field quantity,
$X(\mathbf{r}, t): \mathbb{G} \rightarrow \mathbb{R}$, where $\mathbb{G} = \{1,
\ldots, X\} \times \{1, \ldots, Y\} \times \{1, \ldots, T\}$. As standard, we
denote this as $X(\mathbf{r},t) \in \mathbb{R}^{X \times Y \times T}$.

Loosely speaking, a spatiotemporal system possesses some level of
\emph{organization} if many of its degrees of freedom evolve collectively.
Coherent structures---the form of organization we are most interested
in---implies an emergent higher-level degree of freedom in the system.
Particles in a fluid vortex, for instance, will largely travel together through
a complex fluid flow. Thus, the particles need not be tracked individually; one
can simply track the single vortex \cite{Week96a}. 

\subsubsection{Local Causal States}

This motivates the idea of physics-informed representation learning to extract coherent structures. If a spacetime field $X(\mathbf{r},t) \in \mathbb{R}^{X \times Y \times T}$ is compressed through some encoding, then the encoded representations may leverage the collective organization present in the system. Particularly when representations are learned through an encoder-decoder framework, the collective higher-order degrees of freedom---coherent structures---capture the majority of a system's behavior and so provide a natural set of compressed representations.

A key distinction between linear modal decompositions \cite{tair17a}---such as
POD or the related Dynamic Mode Decomposition \cite{Tu14a}---and the local
causal states is that modal decompositions provide a finite set of
\emph{spatial field templates} (modes). Each spatial mode $\phi(\mathbf{r})$ provides a fixed template with the geometry of a spatial
field $X(\mathbf{r})$. The local causal states, by contrast, provide a finite
set of \emph{localized templates} that are assigned at each individual point in
spacetime. 

The locality of local causal states is achieved through the use of
\emph{lightcones} in the spatiotemporal system. For systems that evolve
according to local interactions, lightcones delineate the causal influence of a
point in spacetime. Specifically, the \emph{past lightcone} $\PLC\stpoint$ of a
spacetime point $\stpoint \in \mathbb{S}$ is the set of all spacetime points in
the past of $\stpoint$ that could possibly influence $\stpoint$ through
propagation of the local interactions: 
\begin{align}
    \PLC\stpoint := \{\stpprime : t'\leq t , ||\site' - \site|| \leq c (t'-t)\}
    ,
\label{eq:plc}
\end{align}
where $c$ is the speed at which local interactions propagate through the system. 
Similarly, the \emph{future lightcone} $\FLC\stpoint$ of a spacetime point
$\stpoint$ is the set of all spacetime points at later times that $\stpoint$
itself can possibly influence through the local interactions, so that:
\begin{equation}
    \FLC(\site, t) := \{(\site', t') : t' > t ||\site'-\site|| \leq c (t'-t)\}
    .
    \label{eq:flc}
\end{equation}
We include the present spacetime point $\stpoint$ as a member of its past lightcone, but not its future lightcone. See Figure~\ref{fig:lightcones}. 

We denote past and future lightcone random variables as $\PLC$ and $\FLC$ respectively, with realizations denoted as $\plc$ and $\flc$. Lightcone realizations $\ell^{\pm}\stpoint$, or \emph{configurations}, are given by assignment of values from a particular spacetime field $X(\mathbf{r},t) \in \mathbb{R}^{X \times Y \times T}$ to spacetime points $\stpoint$ in $\mathtt{L}^{\pm}\stpoint$. 

For systems defined in a discrete spacetime, such as cellular automata
\cite{Rupe18a}, the speed of light $c$ is simply given by the radius of local
interactions. For systems in a continuous spacetime, like fluid flows, $c$ is
given by the speed of sound in the system. As described above in the context of
Lagrangian Coherent Structures, we are interested in organization that emerges
at advective scales in the flow, which are typically much lower velocities than
the speed of sound. In this case, we consider lightcones in the scope of
Lagrangian advection, defined by 
\begin{align}
    \mathbf{y}(t; t_0, \mathbf{y}_0) := F^t_{t_0}(\mathbf{y}_0)
    ~, 
    \label{eq:Lagrangeadvect}
\end{align}
where $\mathbf{y}(t)$ represents the spatial position of a fluid particle or tracer at time $t$. 

Consider a fluid particle $\mathbf{y}$ initially at spacetime point $(\site_0,
t_0)$. The trajectory of the particle $\mathbf{y}(t)$, given by the flow map in
Eq.~(\ref{eq:Lagrangeadvect}), is the solution of the differential equation:
\begin{align}
    \dot{\mathbf{y}} = v\stpoint
    ~,
\end{align}
where $v\stpoint$ is a smooth velocity field. While the velocity field varies
in space and time, we can assign the maximum velocity value $v_{\text{max}}$ so
that $v\stpoint \leq v_{\text{max}}$.

The \emph{Lagrangian lightcones} then are defined by setting $c =
v_{\text{max}}$ in Eqs. (\ref{eq:plc}) and (\ref{eq:flc}). Again,
$v_{\text{max}}$ is typically much smaller than the speed of sound. Lagrangian
lightcones capture local causality derived from Lagrangian advection. A
Lagrangian past lightcone delineates the possible reach of a particle
$\mathbf{y}(t; t_0, \mathbf{y}_0)$ initially at point $(\site_0, t_0)$ through
Lagrangian advection in reverse time. That is, all the spacetime points at
earlier times from where the particle could possibly have come when evolving
according to Lagrangian advection in Eq.~(\ref{eq:Lagrangeadvect}). In
practice, the value of $c$ used can be seen as a hyperparameter that controls
the spatial scale of organization captured by the approximated local causal
states. A larger $c$ captures more coarse-grained structures. 

In addition to delineating causal influence for points in spacetime, note that
lightcones are defined solely in terms of distances in spacetime. This implies
that (i) they are \emph{equivariant} under spacetime isometries, such as
translations and rotations, and (ii) they are well-defined for any spatial geometry with a
distance metric, such as the surface of a sphere or an arbitrary spatial
network \cite{Shal03a}. Thus, they can be used in a wide variety of
spatiotemporal systems, and they transform appropriately under translations,
rotations, and reflections. We will return to this later point shortly. 
Note that planar projections of spherical data, like those we use here for climate data, can break global rotational symmetry, and therefore will not be robust across different projections. However, the local causal states appear to be robust to local rotations in a given projection. For example, the local causal state signature of hurricanes does not change as the hurricane rotates locally. 

Having defined lightcones, we can now define local causal states through the
\emph{local causal equivalence relation}. Two past lightcone configurations are
considered causally equivalent if they have the same conditional distribution
over co-occurring future lightcones: 
\begin{align}
    \plc_i \sim_\epsilon \plc_j &\iff \Pr(\FLC | \PLC=\plc_i) = \Pr(\FLC | \PLC=\plc_j)
    ~.
    \label{eq:causalequiv}
\end{align}
The equivalence classes of the local causal equivalence relation Eq. (\ref{eq:causalequiv}) are the \emph{local causal states}. Individual local causal states are denoted $\xi$ and the set of all local causal states for a given system is $\Xi$. 

A local causal state is a set of past lightcone configurations that all have the same conditional distribution $\Pr(\FLC | \PLC=\plc)$. It can also simply be thought of as the conditional distribution itself, since all past lightcone configurations in a local causal state by definition share that one distribution. 

Here, we consider deterministic systems whose dynamics do not change over time;
e.g., a fluid flow governed by the Navier-Stokes equations with time-independent
parameters. Therefore, the conditional distributions over lightcones $\Pr(\FLC
| \PLC=\plc)$ are also time-independent---a condition known as \emph{conditional stationarity} \cite{cair05a}. For such systems, each past lightcone
configuration $\plc_i$ has a unique and well-defined distribution $\Pr(\FLC |
\PLC=\plc_i)$ and, thus, is also associated with a unique local causal state
defined by that distribution. Therefore, local causal states do not require the system dynamics to be stationary to be well-defined. This allows us to apply local causal states to conditionally-stationary behaviors, like vortex decay in two-dimensional turbulence. 

We define the
\emph{$\epsilon$-function} as the mapping from past lightcone configurations to
their corresponding local causal state $\epsilon: \plc \mapsto \xi$. The
functional form of local causal equivalence relation is given in terms of the
$\epsilon$-function as:
\begin{align}
    \plc_i \sim_\epsilon \plc_j &\iff \epsilon(\plc_i) = \epsilon(\plc_j)
    ~.
    \label{eq:eqsilonmap}
\end{align}

For a given spacetime field $X\stpoint$, each spacetime point $\stpoint \in \mathbb{S}$ has a unique past lightcone configuration $\plc\stpoint$. Applying the $\epsilon$-function then gives a unique local causal state at that point, $\xi \stpoint = \epsilon(\plc \stpoint)$. Therefore, the $\epsilon$-function provides a local point-wise mapping from a spacetime field $X\stpoint$ to an associated \emph{local causal state field}, denoted $S\stpoint = \epsilon(X\stpoint))$. Crucially, the locality of the $\epsilon$-function ensures that a spacetime field $X\stpoint$ and its associated local causal state field $S\stpoint = \epsilon(X\stpoint))$ share the same spacetime coordinate geometry $\mathbb{S}$. For each spacetime point $\stpoint \in \mathbb{S}$, $X\stpoint: \mathbb{S} \rightarrow \mathbb{R}^n$ provides the values of the physical variables at that point and $S\stpoint: \mathbb{S} \rightarrow \Xi$ gives the local causal state $\xi\stpoint = \epsilon(\plc\stpoint))$ at the point. 

And so, as promised, the local causal states are local representations assigned
to each point $\stpoint$ in spacetime $\mathbb{S}$. In this way, the local
causal state field $S\stpoint = \epsilon(X\stpoint))$ provides a
\emph{spacetime segmentation} that assigns a local causal state \texttt{class label} to each point in spacetime. 

The $\epsilon$-function's locality,
together with the equivariance of lightcones under spacetime isometries,
implies that the local causal states are also equivariant under spacetime
isometries. That is, if $g$ is an isometric transformation (preserves spacetime distances) then it commutes with the $\epsilon$-function: $g \circ \epsilon = \epsilon \circ g$. In the context of coherent structure segmentation, this means that local causal states associated with a particular coherent structure,
say hurricanes, do not depend on the hurricane's spacetime location or
orientation. In the Lagrangian Coherent Structure literature, this property
ensures local causal states are an \emph{objective} method for coherent
structure identification \cite{Hall15a,Hadj17a}. 
Analogous to the interpretation of equivariances in neural networks as \emph{geometric deep learning} \cite{bron17a,bron21a}, we can interpret the local causal states as a form of \emph{geometric representation learning} due to their spacetime equivariance and shared coordinate geometry.

Note, too, that locality allows local causal states to capture system structure
across scales. All spacetime points associated to a coherent structure may be
assigned to a set of local causal states uniquely identifying that structure,
independent of the structure's size \cite{Rupe18a}. Larger structures simply
involve a larger number of spacetime points assigned to those structures' states.

Similarly, there is no requirement for balanced statistics across occurrences
of local causal states. For example, in the results shown here there is a often
a single local causal state associated with an ambient ``background'' that
occurs most commonly in spacetime, with local causal states associated to
coherent structures being relatively rarer.

Finally, we emphasize the local causal state approach's flexibility. While all
examples examined here are on discrete Euclidean spacetimes, the local causal
states are well-defined in any spacetime geometry with a distance metric. That
said, analyzing large datasets with two spatial dimensions is already
computationally taxing, even with large supercomputer resources. In principle,
though, the local causal states can be approximated for systems with higher
spatial dimensions. 

\subsubsection{Approximations and Reconstruction}

Consider a \emph{data-generating process} $\mathcal{P}$ that produces spacetime
fields $X \in \mathcal{P}$ and can either be a natural system or a numerical
model. Here, we are interested in real-valued fields, with $X\stpoint \in
\mathbb{R}^n$. Therefore, the spaces of all past and future lightcones for
$\mathcal{P}$, denoted $\mathcal{L}^-$ and $\mathcal{L}^+$ respectively, are
uncountable. Each distribution $\Pr(\FLC | \PLC=\plc)$ is a
continuous probability density. Moreover, they are densities conditioned on
measure-zero events $\PLC = \plc$. There are an uncountable number of past
lightcone configurations $\plc$ and, for each of them, there is a conditional
density $\Pr(\FLC | \PLC=\plc)$ over an uncountable number of future lightcone
configurations.

Thus, in practice approximation is necessary to empirically reconstruct the
conditional distributions $\Pr(\FLC | \PLC=\plc)$. Most importantly, we replace
the measure-zero past lightcone configurations with something of finite
measure. In particular, we partition the space of past lightcones
$\mathcal{L}^-$ into finite-measure events $\past \in \PAlg$ for our empirical
probability estimates. This allows the following empirical densities to be
properly sampled from finite data:
\begin{align}
    \Pr(\FLC | \Past = \past) = \int_{\plc \in \past} \Pr(\FLC | \PLC = \plc) d\mu
    ~,
\end{align}
where $\mu$ is the data-generating distribution over past lightcones. The
finite-measure events partition $\PLCSpace$, so that $\bigcup_i
\past_i = \PLCSpace$, $\past_i \cap \past_j = \emptyset$, and $\mu(\past_i)
\neq 0$ for all $\past_i$.

To achieve such a finite partitioning, we rely on the \emph{continuous histories assumption}~\cite[Assumption 3.1]{Goer12a} which states that if two past lightcone configurations are similar, they should have similar conditional distributions over future lightcones. Formally, this says that the $\epsilon$-function is continuous over the space of past lightcones, and so $\Pr(\FLC | \PLC=\plc_j) \rightarrow \Pr(\FLC | \PLC=\plc_i)$ as $\plc_j \rightarrow \plc_i$. 

The continuous histories assumption leads us to employ distance-based
clustering to create the finite partitioning of the past-lightcone space 
$\PLCSpace$. Let $\gamma : \plc \mapsto \past$ be the mapping from past
lightcone configurations to their distance-based cluster element. We refer to
individual cluster elements $\past$ as \emph{pasts}. The $\gamma$-function
induces an equivalence relation over past lightcone configurations, similar to
the $\epsilon$-function:
\begin{align*}
    \plc_i \sim_{\gamma} \plc_j &\iff \plc_i \in \past_a \text{ and } \plc_j \in \past_a \\
    \plc_i \sim_{\gamma} \plc_j &\iff \gamma(\plc_i) = \gamma(\plc_j)
    ~.
\end{align*}
Two past lightcone configurations are $\gamma$-equivalent if they are assigned to the same distance-based cluster $\past_a$. 

\begin{figure}
\begin{center}
\includegraphics[width=0.6 \columnwidth,trim={3.8cm 1cm 5.8cm 1cm},clip]{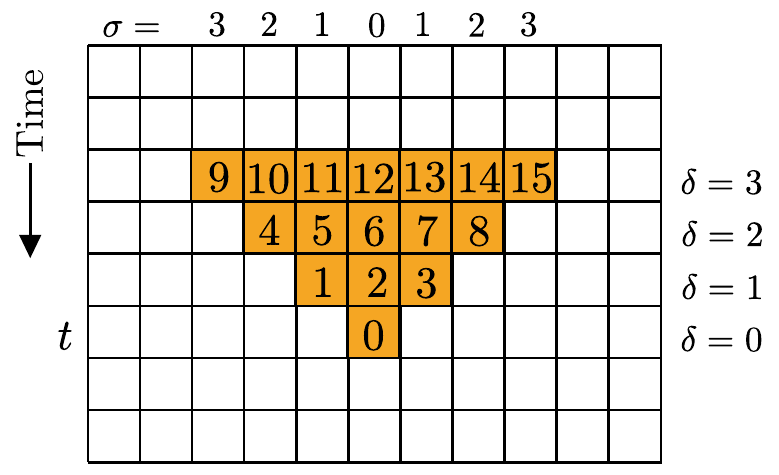}
\end{center}
\caption{Finite vector for the past lightcone configuration $\plc\stpoint$ in
	discrete $1+1$ dimensional spacetime with $c=1$ and finite temporal horizon
	$h^-=3$. Overlaid integers indicate the index of that  site in the
	lightcone vector. $\sigma$ is the space-only internal distance between the
	base of the lightcone, $i=0$, and other sites in each vertical column.
	Similarly, $\delta$ is the time-only internal distance for each site in the
	horizontal row.
	}
\label{fig:plc_vector}
\end{figure}

With distance-based $\gamma$-equivalence, we state the actionable version
of continuous histories our reconstruction algorithm uses as:
\begin{align}
    \gamma(\plc_i) = \gamma(\plc_j) \implies \epsilon(\plc_i) = \epsilon(\plc_j)
    ~.
\end{align}
That is, if two past lightcone configurations are assigned to the same cluster element from a distance-based clustering, we assume them to be ``similar'' and so assume that they have the same distribution over future lightcones. 

To perform distance-based clustering over lightcones, we need a distance metric on the space of lightcones. First, note that in practice lightcones of finite temporal depth must be used. Individual lightcone configurations are collected into a vector, using a canonical ordering of their elements, up to a finite depth \emph{horizon} cutoff $h^{\pm}$ in time. Figure~\ref{fig:plc_vector} shows a past lightcone vector in $1+1$ dimensional spacetime with $c=1$ and $h^- = 3$. Integers overlaid on the lightcone sites indicate the indices of the lightcone vector in the canonical ordering we use. 

Prior work \cite{Jani07a,Goer12a} used a Euclidean distance over lightcones. In practice with a finite horizon cutoff, this means that values in the lightcone below the cutoff are given a uniform weighting when computing distances, while lightcone values beyond the cutoff are given zero weight. To smooth this step discontinuity, our algorithm uses an exponentially-decaying lightcone distance $\mathrm{D}_{\mathrm{lc}}$ given as  
\begin{align}
\mathrm{D}_{\mathrm{lc}}(\mathbf{a}, \mathbf{b}) \equiv \sqrt{(a_0 - b_0)^2 + \ldots + \mathrm{e}^{-\tau d(l)}(a_l - b_l)^2}
~,
\label{eq:lc-dist}
\end{align}
where $\mathbf{a}$ and $\mathbf{b}$ are length-$l$ finite lightcone vectors, as depicted in Figure~\ref{fig:plc_vector}, and $\tau$ is the decay rate. 

The decay is applied relative to an internal distance $d(i)$ between the base
of the lightcone vector, with index $i=0$, and other sites in the lightcone
with indices $0 < i \leq l$. Figure~\ref{fig:plc_vector} shows a space-only
internal distance as $d(i) = \sigma(i)$ and a time-only internal distance as
$d(i) = \delta(i)$. An internal spacetime distance is given as $d(i) =
\sqrt{\sigma(i)^2 + \delta(i)^2}$. Results here use the latter spacetime
internal distance. This family of exponentially-decaying lightcone distances
smooth the step-discontinuity of a simple Euclidean distance by giving less
significance to sites in the lightcone that are further away from the present
(the base of lightcones). 

In addition, the exponential decay has a practical benefit of reducing the
effective dimensionality of the lightcone vectors during clustering, as
discussed in more detail in Ref.~\cite{Rupe19a}. We use K-Means \cite{kmeans}
for our distance-based clustering of lightcones. Comparison with DBSCAN
\cite{dbscan}, along with the distributed High-Performance Computing (HPC)
implementations of these algorithms, is also given in Ref.~\cite{Rupe19a}. 

\begin{figure*}
\begin{center}
\includegraphics[width=1.0 \textwidth]{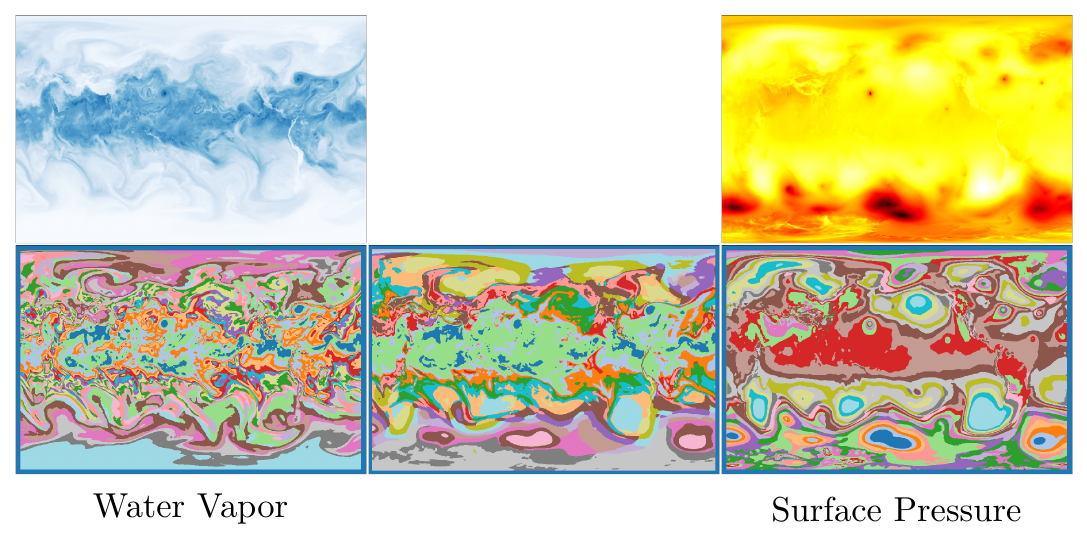}
\end{center}
\caption{Local causal state segmentations of TMQ and PSL, with a linear
	multivariate combined segmentation in between.
	}
\label{fig:tmq-psl}
\end{figure*}

After performing a $\gamma$-partitioning of $\PLCSpace$ using K-Means
clustering, the densities $\Pr(\FLC | \Past = \past) = \int_{\plc \in \past}
\Pr(\FLC | \PLC = \plc) d\mu$ can be empirically sampled from finite data. To
simplify further for our HPC implementation, we also perform a (separate)
K-Means clustering over future lightcones to similarly produce finite-measure
\emph{futures} $\future_i$. This gives a discrete approximation $\Pr(\Future |
\Past=\past_i)$ that can be sampled by simple counting. That is, each
co-occurring set of pasts $\past_a = \gamma^-(\plc\stpoint)$ and futures
$\future_b = \gamma^+(\flc\stpoint)$ are collected in a matrix $\mathfrak{P}$,
where $\mathfrak{P}_{ab}$ is the number of co-occurrences of $\past_a$ and
$\future_b$. 

With the empirical distributions $\Pr(\Future | \Past=\past)$ we use an
empirical approximation to causal equivalence over pasts, rather than over past
lightcone configurations. (In principle, the densities $\Pr(\FLC |
\Past=\past)$ can also be used.) If two empirical future distributions are
close, according to some empirical test, then their pasts are $\psi$-equivalent:
\begin{align*}
\past_i \sim_\psi \past_j &\iff \Pr(\Future | \past_i) \approx \Pr(\Future | \past_j)\\
\past_i \sim_\psi \past_j &\iff \psi(\past_i) = \psi(\past_j)
~.
\end{align*}
We use hierarchical agglomerative clustering with a chi-squared similarity test ($p=0.05$) for $\psi$-equivalence. 

Thus, the approximation of the $\epsilon$-function our reconstruction
algorithm uses is:
\begin{align}
    \epsilon(\plc) \approx \psi\bigl(\gamma(\plc)\bigr)
~.
\label{eq:epsilonapprox}
\end{align}

Note that because past lightcones of depth $0$ reduce
to the lightcone base---simply the field value at that point in spacetime---our
$\gamma$ clustering over depth $0$ past lightcones reduces to a standard
K-Means clustering over the spacetime field. Recent insights into
deep learning utilizing maximum affine spline operators \cite{bale18a} shows
that trained deep learning models partition their input space, just as under
distance-based clustering. This implies ``that a D(eep)N(eural network)
constructs a set of signal-dependent, class-specific templates''
\cite{bale18a}, formally similar to local causal state representations. Whereas
a neural network uses ``ground truth'' labels during the supervised learning
process to build its templates, the local causal states must be learned in an
unsupervised fashion. Our use of lightcones is motivated by weak causality and
a local spacetime generalization of delay-coordinate embeddings and their
associated intrinsic geometry \cite{Pack80}. This, combined with predictive
equivalence, provides the necessary physics of organization \cite{rupe22a} with
which to extract coherent structures \cite{Rupe18a}.

\subsubsection{Multivariate Interpolation}
To reconstruct approximate local causal states for multivariate systems with
$m$ physical fields, we aggregate lightcone distances over all the fields.
The simplest way to do this is to use the following tensor lightcone metric:
\begin{align*}
\mathrm{D}_{\ell}(\mathbf{a}, \mathbf{b}) \equiv \sqrt{w_0\biggl((a_0^0 - b_0^0)^2 + \ldots + \mathrm{e}^{-\tau d(n)}(a_n^0 - b_n^0)^2\biggr) + \ldots + w_m\biggl((a_0^m - b_0^m)^2 + \ldots + \mathrm{e}^{-\tau d(n)}(a_n^m - b_n^m)^2\biggr)}
~.
\end{align*}

A practical advantage of this particular metric is that it can be achieved by simply concatenating the scalar lightcone vectors taken from each field. Empirically, we observe that this metric yields a local causal state field that is a weighted interpolation of the component fields, with relative weighting given by the coefficients $w_i$. We use the water vapor and surface pressure fields of the CAM5.1 model to demonstrate this. 

From visual inspection, signatures of hurricanes can be seen in the column-integrated water vapor field (TMQ) and surface pressure field (PSL), among others. Running local causal state segmentation on these fields alone is not sufficient for producing a unique segmentation class corresponding to hurricanes. The segmentation class of the TMQ field corresponding to hurricanes also shows up in the tropics in regions of high water concentration, which are not hurricanes. Similarly, the PSL segmentation class corresponding to hurricanes shows up in similar, but larger scale, pressure patterns in the extra tropics (particularly in the southern hemisphere). 

Figure~\ref{fig:tmq-psl} gives example segmentations of these fields. The
middle of Figure~\ref{fig:tmq-psl} shows a multivariate segmentation
incorporating both fields, with the PSL field given $\frac{1}{4}$ the weight of
TMQ in the lightcone tensor metric described above. Qualitatively, one sees how
clustering with this metric produces a weighted interpolation between
segmentations of the two fields separately. In this case, the tropics and
northern hemisphere are more reflective of the TMQ field, while the southern
hemisphere is more reflective of the PSL field. Unfortunately, interpolating
features from both fields together does not result in the desired unique
hurricane segmentation class.

\subsection{Application Problems and Data Sets}
\subsubsection{Turbulence}

The formal study of emergent organization had its genesis with B\'{e}nard's
work \cite{Bena01a} on the spontaneous formation of fluid convection cells
\cite{Rayl16a,Chan68a,Buss78a,Fens79a,Stei85} at the turn of the
20$^{\text{th}}$ century. Around the same time, the dawn of quantum mechanics
took prominence in fundamental physics. Even up to the present, however, fluid
turbulence has remained lingering as ``the last mystery in classical physics''.
Arguably, the primary ``mystery'' is that of emergent organization
\cite{Cros93a}, with turbulence being the flagship instantiation
\cite{Heis67a}. 

The importance of secondary large-scale coherent structures in general theories
of turbulence was recognized in the latter half of the 20$^{\text{th}}$
century, as documented in Ref.~\cite{liu88a}. The first exploration of
turbulent coherent structures were statistical in nature. It was observed that
turbulence has a statistical separation of scales in fluctuations about the
mean flow. There are smaller, fine-scaled ``random'' fluctuations on top of
larger ``not random'' fluctuations, the latter being identified as coherent
structures \cite{liu88a}. This statistical organization was identified and
corroborated through characteristic signatures in the correlation function
\cite{liep52a,town56a}. 

Structural approaches, rather than statistical, appeared later, particularly with the introduction of the Proper Orthogonal Decomposition in the study of turbulence \cite{luml67a}. The POD approach follows an Eulerian paradigm, viewing the full spatial fields as the state of a spatially-extended dynamical systems. The spatial fields are then decomposed into POD modes, which have an optimality property that the leading modes capture the dominant contributions to kinetic energy. A (relatively) low-dimensional dynamical system analysis of the evolution of the POD modes provides an analysis of the flow in terms of its dominant \emph{energetic} coherent structures \cite{Holm12a}. 

Also building on ideas and tools from low-dimensional dynamical systems, the geometric theory of coherent structures follows the Lagrangian paradigm. Rather than seeking to capture energetically-dominant structures, the Lagrangian approach seeks coherent structures that organize advective transport in the flow. Lagrangian coherent structures are thus found as the most attracting, repelling, or shearing material surface in the flow \cite{Hall15a}. 

The local causal states are a general representation learning method like POD.
However, when applied to fluid flows using Lagrangian lightcones, as described
above, they are more closely related to Lagrangian coherent structure
approaches. 

To analyze the ability of local causal states to capture fluid coherent
structures we examine vortex dynamics in two-dimensional turbulence. This
particular flow is ideal for several reasons. First, it is two-dimensional,
making it easier to analyze with computationally expensive algorithms. However,
unlike simple two-dimensional flows like von K\'{a}rman vortex streets, the
two-dimensional turbulent flow supports behaviors characteristic of anisotropic
turbulence in three dimensions, particularly in geophysical flows
\cite{mcwil83a}.

Second, the coherent structures that emerge in two-dimensional
turbulence---coherent vortices---are paradigmatic fluid coherent structures.
Moreover, while a general theory of the emergent vortex dynamics remains
elusive, the dynamics are known empirically and are qualitatively simple. Two
like-signed vortices, under the right conditions, will pairwise merge when
they are close enough. Third, this pairwise merging behavior results in a
power-law decay of the total number of vortices in the flow over time
\cite{mcwil90a,carn91a}. Identification of the power-law decay provides a
quantitative metric to test potential vortex coherent structure detection
methods. Note that the identification of the ``theoretical'' range for the decay rate $\nu = -(0.71-0.75)$ is identified using a specialized vortex identification algorithm based on vorticity thresholding along with geometric considerations \cite{mcwil90a}.

We use data generated from the Fluid2D solver publicly available on GitHub (\url{https://github.com/pvthinker/Fluid2d/tree/master/experiments/Twodim_turbulence}). The local causal state segmentation used to count vortex cores, shown in Figure~\ref{fig:vortex_decays} (b), uses the following parameters: past lightcone horizon $14$, future lightcone horizon $2$, propagation speed $c=1$, $K=3$ for past lightcone K-Means, $K=25$ for future lightcone K-Means, and a spacetime decay with decay rate $\tau=0.25$ for the lightcone distance metric. 

\begin{figure*}
\begin{center}
\hspace*{0.0cm}\includegraphics[width=0.7 \textwidth]{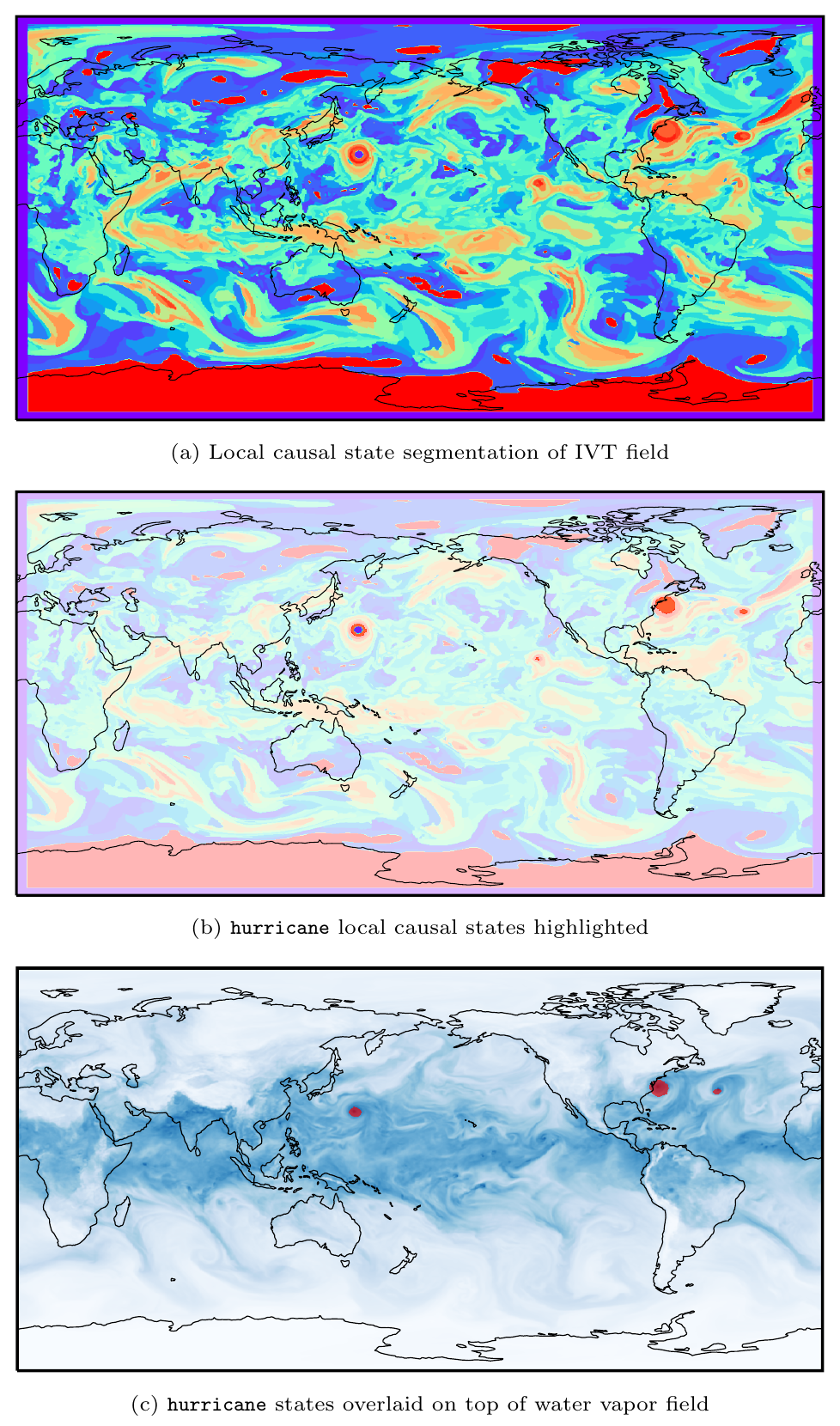}
\end{center}
\caption{Creation of hurricane tracker using local causal state segmentation of the IVT field.
	}
\label{fig:ivt_hurricane_mask}
\end{figure*}

\subsubsection{Climate}
While spacecraft allow for direct video observations of portions of the
atmosphere, similar to that of Jupiter shown below, direct observations of the
relevant physical fields of the full atmosphere are not possible. Numerical
models are used to aggregate inhomogeneous data sources and create \emph{data
images} of the physical fields most consistent with the observations
\cite{edwa10a}. This process is known as \emph{reanalysis} and represents the
closest one can come to ``true'' observational climate data \cite{park16a}. 

In contrast to reanalysis data, general circulation models (GCM) simulate the
dynamics of the atmosphere to make short-term predictions (e.g., numerical
weather prediction) and long-term climate forecasts. Long-term forecasts are
the key tool for analyzing the potential future impacts of global warming
\cite{edwa10a,eyri16a}. Tools such as the local causal states provide an
enhanced level of analysis to answer detailed questions, such as how the
intensity and dynamics of extreme weather events will change under different
warming scenarios. 

In this study, we use data from the historical calibration run of the high resolution $0.25^\circ$ CAM5.1 GCM~\cite{wehn14a}. Single-precision climate variables are stored on an 1152 x 768 spatial grid, with a temporal resolution of 3
hours. The segmentation results shown here use data from the autumn of 2015 of
the CAM5.1 historical run $1$. The physical fields used for the various local
causal state segmentation results are the column-integrated water vapor (TMQ),
the near-surface velocity components (U850) and (V850), as well as the
integrated vapor transport (IVT) field, defined as:
\begin{align*}
    \bigg[\biggl(\frac{1}{g}\int_{1000hPa}^{600hPa} \; qu \;dp\biggr)^2 + \biggl(\frac{1}{g}\int_{1000hPa}^{600hPa} \; qu \;dp\biggr)^2 \bigg]^{1/2}
    ~,
\end{align*}
where here $q$ is water vapor, $u$ and $v$ are wind velocity components, and
$p$ is pressure.

\subsubsection{Hurricane Tracker}

We create a hurricane tracker using a univariate local causal state segmentation of the IVT field, as shown in Figure~\ref{fig:ivt_hurricane_mask}. A snapshot of the local causal state field $S \stpoint$ from this segmentation is shown in (a). From visual inspection, we find a particular set of \texttt{hurricane} states that seem to co-occur with hurricanes. That is, it appears that (almost) all spacetime points $\stpoint$ such that $S\stpoint$ is one of the \texttt{hurricane} states are points where a hurricane is present (as seen in the TMQ field). The \texttt{hurricane} states are highlighted in the local causal state segmentation of Figure~\ref{fig:ivt_hurricane_mask} (b). To aid the visual identification of the \texttt{hurricane} states with actual hurricanes, a snapshot of the water vapor field (TMQ) is shown in (c) with spacetime points $\stpoint$, such that $S\stpoint$ is one of the \texttt{hurricane} states, given a red mask overlaid on top of the vapor field. We emphasize again that this is possible due to the shared coordinate geometry of the physical fields and the local causal state field. 

The local causal state segmentation of the IVT field, shown in Figure~\ref{fig:ivt_hurricane_mask}, was created using the following parameters: past lightcone horizon $16$, future lightcone horizon $3$, propagation speed $c=1$ (one spatial grid cell per single $3$-hour time step), $K=24$ for past lightcone K-Means, $K=50$ for future lightcone K-Means, and decay rate $\tau=1.0$ for the lightcone distance metric. 

\subsubsection{General Extreme Weather Events}

The more general extreme weather local causal states shown in Figure~\ref{fig:EWE-seg} (a) are produced using a multivariate segmentation that combines the TMQ, U850, and V850 vapor and velocity fields using a tensor lightcone metric based on IVT,
\begin{align}
\mathrm{D}_{\mathrm{lc}}(\mathbf{a}, \mathbf{b}) \equiv \sqrt{\biggl((a_0^q a_0^u - b_0^q b_0^u)^2 + \ldots + \mathrm{e}^{-\tau d(n)}(a_n^q a_n^u - b_n^q b_n^u)^2\biggr) + \biggl((a_0^q a_0^v - b_0^q b_0^v)^2 + \ldots + \mathrm{e}^{-\tau d(n)}(a_n^q a_n^v - b_n^q b_n^v)^2\biggr)}
~.
\label{eq:ivt_alt_metric}
\end{align}
Here, $\mathbf{a}$ and $\mathbf{b}$ are lightcone tensors such that lower indices are the lightcone locations (as shown in Figure~\ref{fig:plc_vector}) and upper indices are the physical fields; $q$ is water vapor, and $u$ and $v$ are the near-surface velocity components. Using this metric, the local causal state segmentation shown in Figure~\ref{fig:EWE-seg} (a) is reconstructed using the following parameters: past lightcone horizon $6$, future lightcone horizon $3$, propagation speed $c=2$, $K=14$ for past lightcone K-Means, $K=20$ for future lightcone K-Means, and spatiotemporal decay rate $\tau = 0.5$. 

As described above, the non-white states in Figure~\ref{fig:EWE-seg} (a) represent general coherent structures, according to the local causal state definition \cite{Rupe18a}, that we hypothesize correspond to extreme weather events (EWEs). These include known structures like hurricanes and atmospheric rivers, as well as additional as-of-yet unknown structures. To test the relevance of these structures (particularly the unknown ones) to weather extremes, we examine their relation to precipitation extremes using the PRECT 3 hour cumulative precipitation field of the CAM5.1 model. Note that PRECT is not used to reconstruct the \texttt{EWE} local causal states. 

We follow a similar procedure as Reference \cite{dagon22a}, which uses supervised deep learning to track fronts and associate precipitation extremes with the identified fronts. Here, we count the number of global precipitation extremes that co-occur with the \texttt{EWE} local causal states (the non-white states in Figure~\ref{fig:EWE-seg} (a)). That is, we find the spacetime locations $\stpoint$ of all extreme precipitation events of a given percentile and check to see if $S \stpoint$ is one of the \texttt{EWE} states. If so, the extreme precipitation event co-occurs with the \texttt{EWE} local causal states. Results are shown above in Table~\ref{tab:extremes}. 



\subsubsection{TECA Details}

TECA \cite{Prab12a} is a software package that provides a unified platform for
extreme weather and climate analytics. The results shown in this work use two
distinct heuristics, one for hurricanes and one for ARs, both implemented in
TECA. 

TECA hurricane segmentation follows the TSTORMS code, originally developed by
the Geophysical Fluid Dynamics Laboratory and described in Ref. \cite{Knut07a}.
In this, explicit physical thresholds are set to determine hurricanes as local
maxima of vorticity and temperature, and local minima of pressure. Implemented
with the command line function \texttt{teca\_tc\_detect}, this algorithm
identifies hurricane centers. We use the default settings for this function. 

Complete hurricane segmentation is achieved by including wind radius
information \cite{Chava15a}. This uses wind radial profiles that define
category strengths of hurricanes. We wish to be as unrestrictve as possible in
identifying possible EWEs with TECA, so we use the ``category 0'' wind profile,
which gives the largest storm radius, provided by the command line function
\texttt{teca\_tc\_wind\_radii}. That is, each hurricane center provided by
\texttt{teca\_tc\_detect} has an associated ``cateogry 0'' wind radius given by
\texttt{teca\_tc\_wind\_radii} that provides the size of the category 0
hurricane at each time. This common hurricane segmentation approach necessarily
gives hurricanes with circular shapes, whereas the local causal states do not
have such explicit shape constraints.

\begin{figure*}
\begin{center}
\hspace*{0.0cm}\includegraphics[width=0.8 \textwidth]{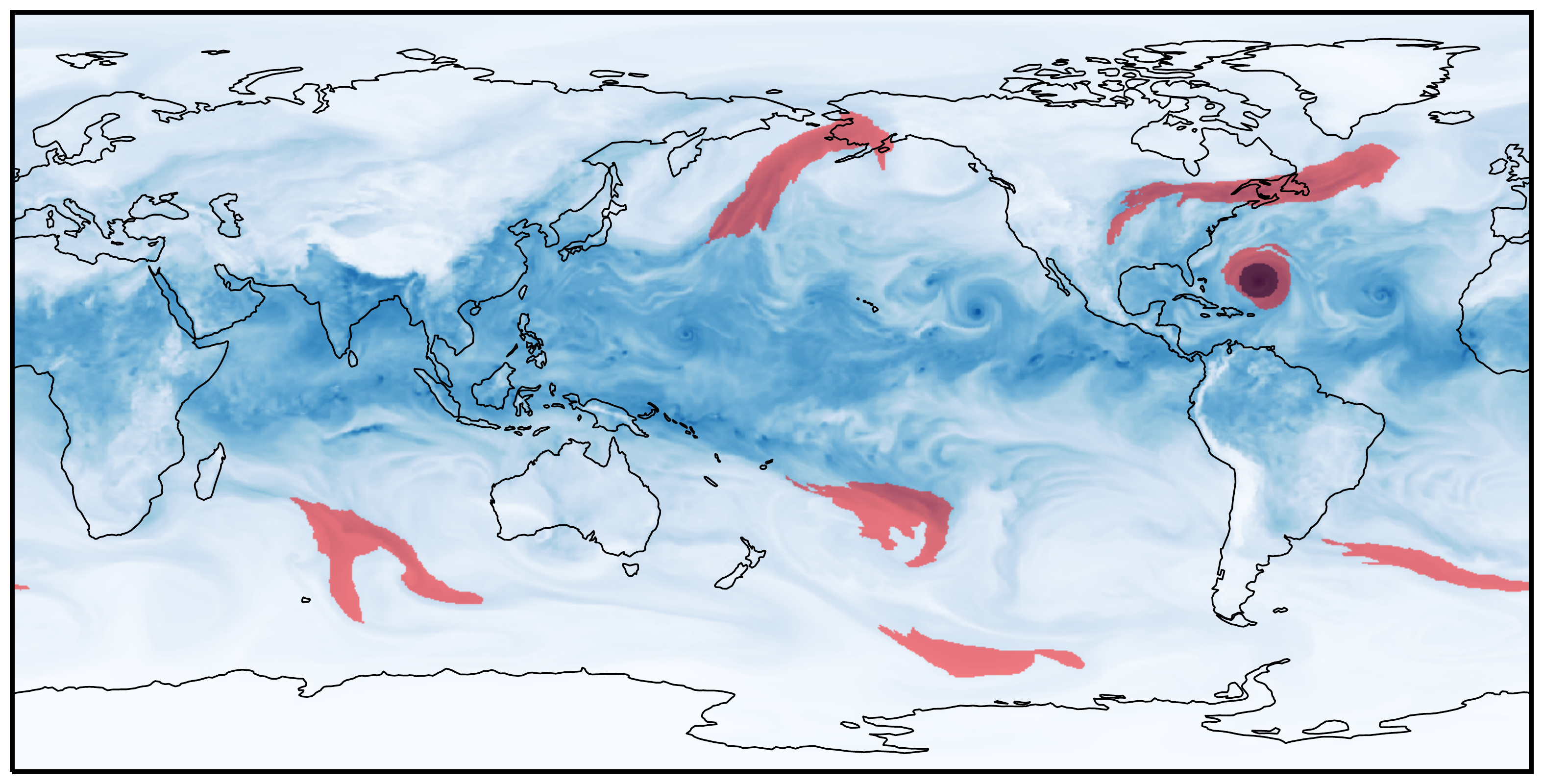}
\end{center}
\caption{TECA AR heuristic false identification of a hurricane as an AR.
	}
\label{fig:teca-false}
\end{figure*}

AR segmentation is performed separately from hurricane segmentation using the
TECA-BARD algorithm \cite{Obrien20a}. In this, a standard AR heuristic is first
employed that detects candidate ARs from the integrated vapor transport (IVT)
field based on threshold, size, and location. A Bayesian framework is then
applied to tune the specific parameters (e.g., the hard threshold) of this
heuristic to best match the AR segmentation from a small dataset hand-labeled
by climate experts \cite{prab21a}. The output of TECA-BARD is an AR likelihood,
which gives the probability of the presence of an AR at each spacetime point.
To create a binary AR segmentation with ARs either present or not, we use a
likelihood threshold of $2/3$. 

Since heuristics such as TECA are not ground-truth, they may produce false
event detections. A clear example is shown in Figure~\ref{fig:teca-false}, in
which TECA-BARD misidentifies a hurricane off the east coast of North America
as an AR. The light red colors in Figure~\ref{fig:teca-false} are \texttt{AR}
segmentation labels from TECA-BARD. The darker-colored circle is the distinct
\texttt{hurricane} TECA segmentation, again performed entirely separately from
TECA-BARD. Thus, spacetime points inside the dark circle are labeled as
\texttt{hurricane} by the TECA hurricane heuristic and also labeled as
\texttt{AR} by the TECA-BARD AR heuristic.

The TECA codebase is developed and maintained by LBNL and is publicly available on GitHub: \url{https://github.com/LBL-EESA/TECA}.

\begin{figure*}
\begin{center}
\hspace*{-1.0cm}\includegraphics[width=0.7 \textwidth]{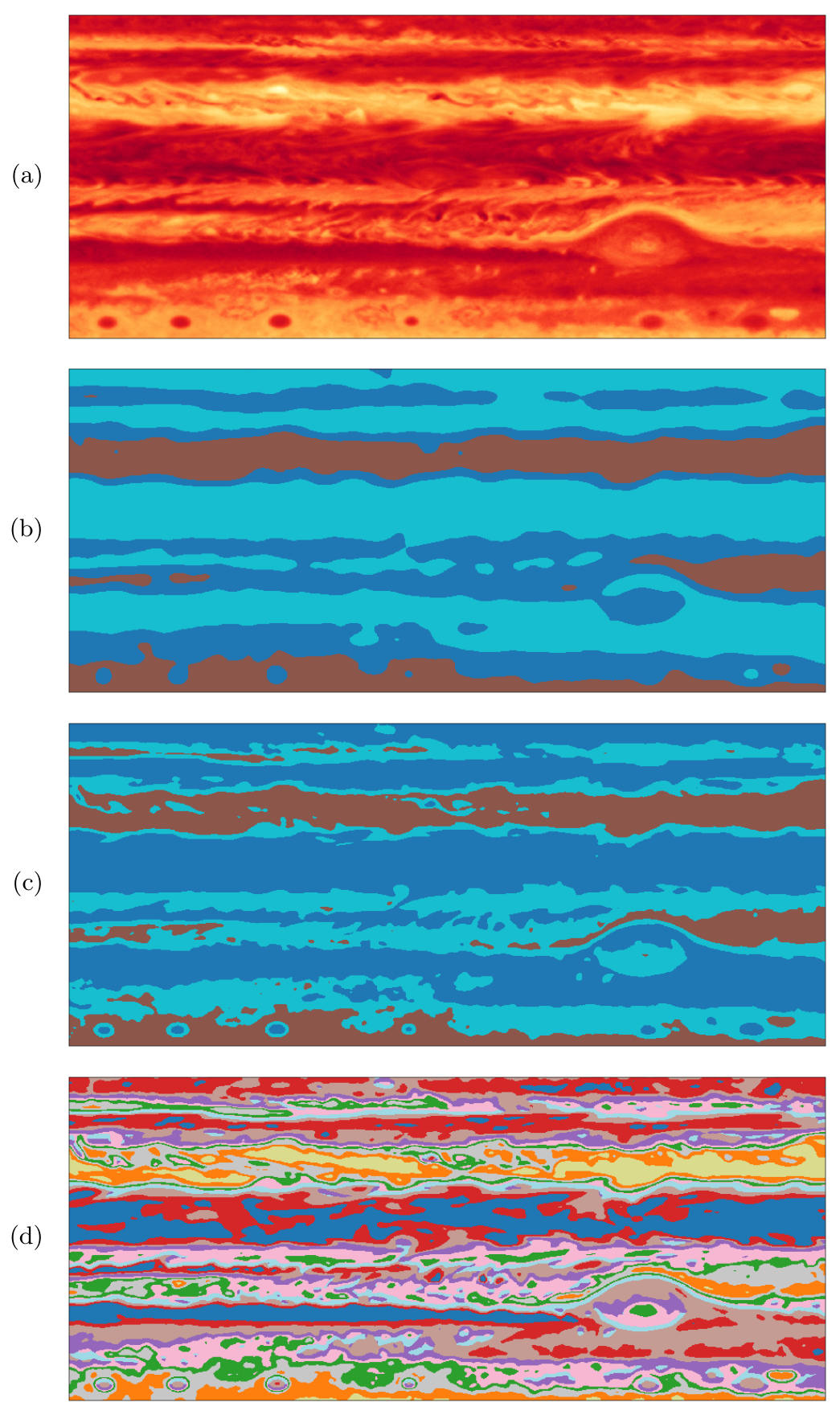}
\end{center}
\caption{Suite of local causal state segmentations (b)-(d) of Jupiter's
	clouds (a).
	}
\label{fig:jupiter}
\end{figure*}

\subsection{Complexity and Symmetry Breaking: \\Clouds of Jupiter}

We close by examining several local causal state segmentations of the clouds of
Jupiter. These examples compare the effects of various inference parameters on
the resulting segmentation and serve to illustrate the idea of increasing
complexity through symmetry breaking.

Figure~\ref{fig:jupiter} shows the results with the observable grayscale image
of Jupiter's clouds, taken from the NASA Cassini spacecraft \cite{cassini},
shown in (a). The largest scale structures that appear are the east-west zonal
bands separated by strong jet streams \cite{Hadj16a}. To capture this large
scale structure, we first perform a segmentation using a (relatively) large
speed of light for the lightcones $c=4$. (Memory requirements of the algorithm
scale most rapidly with $c$). This effectively provides a larger spatial
convolution kernel with each lightcone and thus averages over the smaller scale
degrees of freedom within that kernel. The result, shown in (b), provides
segmentation classes that follow the large scale structure and zonal bands. In
addition, the boundaries between classes corresponding to the zonal bands
identify the east-west jet streams that separate the bands. 

To capture more smaller-scale detail, particularly at the boundaries of zonal
bands, the segmentation in (c) uses the exact same parameters as (b), but
includes a nonzero spatial decay in its lightcone metric. This decay
effectively reduces the size of the spatial convolution kernels and so captures
smaller scale detail. Most noticeable is the appearance of the ``string of
pearls'' in the southern hemisphere that are much more clearly outlined in (c)
than in (b). Note that the different spatial scales found between (b) and (c)
could not be recovered using a standard K-Means segmentation; K is set to $3$
for both cases, only the spatial decay rate differs. 

While decreasing the (effective) size of the spatial kernel reveals smaller
scale details at zonal band boundaries, we can see from (a) that there is
additional turbulent structure within the bands themselves. The segmentation in
(d) reveals more of this internal structure by decreasing the size of the
spatial kernel further using $c=1$ and increasing the number of segmentation
classes found; i.e., increasing $K$ in K-Means. The increased complexity
corresponds to system's broken symmetries, reminiscent of the bifurcation
theory of pattern formation. That is, points within a zonal band belong to the
same segmentation class in (b), signaling they are part of the same large scale
structure. However, they may correspond to different classes in (d),
identifying the points belong to different structures at smaller scales. The
separation of scales then represents an elevated level of complexity in the
system's spatial structure---a coexistence of organization and chaos
(turbulence).

Note that in all segmentation analyses the local causal states vary more in the
vertical direction of the images than the horizontal. As mentioned, the
east-west horizontal bands are the most prominent structures in Jupiter's
atmosphere, created by strong jet streams that act as transport barriers. We
again emphasize that Lagrangian Coherent Structures and our local causal state
approach using Lagrangian lightcones discover coherent structures associated
with material transport. It is not surprising then that local causal states are
extended in the east-west direction following the zonal banding. Particularly
evident in the coarse-grained segmentation shown in (b), the north-south
boundaries between two local causal state bands then indicate the presence of a
jet that acts as a transport boundary between the bands. 

\section{Code Availability}
Supporting Python source code, SLURM run scripts, parameter logs, and Jupyter notebooks displaying results and figures can be found at \url{https://github.com/adamrupe/Emergent-Organization}. 

\section{Author Contributions}
AR and JPC developed the theoretical framework; KK, AR, and JPC conceptualized
the problem applications; AR and KK performed the application experiments and
analysis; AR wrote the prototype code; NK and AR wrote the distributed HPC
code; AR and JPC wrote the manuscript.

\section{Acknowledgments} 
The authors thank Vladislav Epifanov, Oleksandr Pavlyk, Frank Schlimbach, Mostofa Patwary, Sergey Maidanov, and Victor Lee for their help in developing the HPC implementation of local causal state reconstruction. We thank Nicolas Brodu, Jian Lu, Anastasiya Salova, and Mikhael Semaan for helpful comments and feedback. 
We also thank Michael Wehner for provding the IVT fields of the CAM5.1 data, Burlen Loring for help implementing TECA hurricane segmentation, Travis O'Brien for sharing TECA AR segmentation, Wahid Bhimji for help with NERSC resources, and Prabhat initiating and leading the collaboration Project DisCo that ultimately led to the results in this work. 

Part of this research was performed while AR was visiting the Institute for
Pure and Applied Mathematics, which is supported by the National Science
Foundation grant DMS-1440415. AR acknowledges the support of the U.S.
Department of Energy through the LANL/LDRD Program and the Center for Nonlinear
Studies. JPC would like to acknowledge Intel\textsuperscript{\textregistered}
for supporting the Intel Parallel Computing Center at UC Davis. KK was
supported by the Intel\textsuperscript{\textregistered} Big Data Center. This
research is based upon work supported by, or in part by, the U. S. Army
Research Laboratory and the U. S. Army Research Office under contracts
W911NF-13-1-0390 and W911NF-18-1-0028, and the U.S Department of Energy (DOE),
Office of Science, Office of Biological and Environmental Research,
\textbf{Earth and Environmental Systems Modeling program}. This work used
resources of the National Energy Research Scientific Computing Center, a DOE
Office of Science User Facility supported by the Office of Science of the U.S.
Department of Energy under Contract No. DE-AC02-05CH11231. The Pacific
Northwest National Laboratory (PNNL) is operated for DOE by Battelle Memorial
Institute under contract DE-AC05-76RLO1830. 

\end{document}